# A Survey of Social Cybersecurity: Techniques for Attack Detection, Evaluations, Challenges, and Future Prospects

Aos Mulahuwaish[a], Basheer Qolomany[b], Kevin Gyorick[a], Jacques Bou Abdo[c], Mohammed Aledhari[d], Junaid Qadir[e], Kathleen Carley[f] and Ala Al-Fuqaha[g]

[a] Department of Computer Science and Information Systems, Saginaw Valley State University, University Center, 48710, USA
[b] Department of Medicine, College of Medicine, Howard University, Washington, DC, 20059, USA
[c] School of Information Technology, University of Cincinnati, Cincinnati, OH 45221, USA
[d] Department of Data Science, University of North Texas, Denton, TX, 76207, USA
[e] Computer Science and Engineering Department, Qatar University, Doha, Qatar
[f] School of Computer Science, Carnegie Mellon University, 5000 Forbes Ave, Pittsburgh, PA, 15213, USA
[g] Information and Computing Technologies (ICT) Division, College of Science and Engineering (CSE), Hamad Bin Khalifa University, Doha, Qatar

## ARTICLE INFO

## ABSTRACT

In today's age of digital interconnectedness, understanding and addressing the nuances of social cybersecurity have become paramount. Unlike its broader counterparts, information security and cybersecurity, which are focused on safeguarding all forms of sensitive data and digital systems, social cybersecurity places its emphasis on the human and social dimensions of cyber threats. This field is uniquely positioned to address issues such as different social cybersecurity attacks like cyberbullying, cybercrime, spam, terrorist activities, and community detection. The significance of detection methods in social cybersecurity is underscored by the need for timely and proactive responses to these threats. In this comprehensive review, we delve into various techniques, attacks, challenges, potential solutions, and trends within the realm of detecting social cybersecurity attacks. Additionally, we explore the potential of readily available public datasets and tools that could expedite research in this vital domain. Our objective is not only to tackle the existing challenges but also to illuminate potential pathways for future exploration. Through this survey, our primary focus is to provide valuable insights into the rapidly evolving landscape of social cybersecurity. By doing so, we aim to assist researchers and practitioners in developing effective prediction models, enhancing defense strategies, and ultimately fostering a safer digital environment.

## 1. Introduction

In today's digital era, the Internet, especially social media platforms, plays a significant role in shaping public opinions, attitudes, and beliefs. Unfortunately, the credibility of scientific information sources is often undermined by the spread of misinformation through various means, including technology-driven tools like bots, cyborgs, trolls, sock-puppets, and deep fakes. This manipulation of public discourse serves antagonistic business agendas and compromises civil society. In response to this challenge, a new scientific discipline has emerged: social cybersecurity.

**What is Social Cybersecurity?**

"*People often represent the weakest link in the security chain and are chronically responsible for the failure of security systems*."—Bruce Schneier [1].

Social cybersecurity is a field centered around human factors [2; 3], striving to protect individuals and organizations from threats emanating from social manipulation and malicious digital activities [4; 5]. As emphasized by the epigraph of this section, humans are frequently considered the most vulnerable element in the security chain, with numerous underlying reasons for this susceptibility. Schneier [1; 6] delineates six principal factors attributing to human vulnerability:

1. The inherent human struggle with effective risk perception and analysis.
2. The inadequacy of human intuition for processing infrequent events.
3. The perilous overreliance and unwarranted trust people place in computers.
4. The impractical expectation for consistent, intelligent security decisions by everyone.
5. The frequent origination of security breaches from malicious insiders.
6. The commonplace and seemingly effortless occurrences of security lapses due to social engineering.

Social cybersecurity merges applied research in computational science with computational social science techniques to discern, counter, and understand threats associated with social communication. It is an interdisciplinary domain synthesizing elements of high-dimensional network analysis, data science, machine learning (ML), natural language processing (NLP), and agent-based simulation. By utilizing these tools, crucial insights about social media and internet users are unearthed, enabling comprehension of their tactics and fostering the formulation of robust counterstrategies against social manipulation.

As social cybersecurity continues to evolve, new forms of digital manipulation are emerging. One example is the rise of virtual influencers and their impact on digital trust.

Research by Hong et al. (2024) explores how identity and race in virtual influencer interactions influence public perceptions, which is crucial for understanding the dynamics of social cybersecurity threats, particularly in misinformation campaigns [7].

Moreover, tools in social cybersecurity are pivotal in evaluating and foreseeing the repercussions of influence operations on social media, thereby reinforcing the security of online social interactions and diminishing the adverse effects of malicious influence. These insights are invaluable, enhancing intelligence and contributing to advancements in forensic research practices.

**Historical Developments**

The field of social cybersecurity has evolved significantly over the past two decades, emerging at the intersection of cybersecurity, psychology, and social network analysis. Early research in the 2000s focused on integrating social network analysis (SNA) to understand the spread of cyber threats and the role of human behavior in security vulnerabilities [8]. By the 2010s, increased attention was given to the human factors influencing cybersecurity, leading to advancements in social engineering defenses and misinformation detection [9]. The rise of deepfake technology and AI-generated disinformation in recent years has further expanded the field, necessitating AI-driven countermeasures using machine learning (ML), natural language processing (NLP), and dynamic network analysis (DNA) [10]. These technological advancements have not only enhanced cyber defense strategies but also introduced new challenges, as adversaries increasingly exploit generative AI and automation for more sophisticated attacks [11]. Looking ahead, the field will likely continue to evolve with interdisciplinary approaches, integrating insights from behavioral science, digital forensics, and forensic cyberpsychology to mitigate emerging cyber threats effectively [12].

**What Sets Social Cybersecurity Apart?**

It is essential to differentiate social cybersecurity from related fields, such as information security [13; 14] and cybersecurity [15; 16]. Information security is a broad discipline aimed at safeguarding all sensitive data, encompassing both digital and physical aspects. In contrast, cybersecurity, which is a subset of information security, focuses specifically on protecting computer systems, networks, and digital data. Social cybersecurity, on the other hand, places a unique emphasis on the human dimension within the realm of cybersecurity. It delves into the relationship between human behavior, social interactions, and cybersecurity threats. Despite their distinct focuses, these three fields share a common overarching goal: the comprehensive protection of data and systems. Together, they contribute to a multi-faceted, robust approach to security in our increasingly interconnected world, as illustrated in Figure 1. Furthermore, there is a differentiation in cognitive security [17; 18], which centers on the psychological aspects of cybersecurity. In contrast, social cybersecurity explores societal dynamics and investigates the effects of cyber threats on social structures, narratives, and societal trust in digital technologies and institutions. Each of these fields collectively strives for digital security, providing unique perspectives and approaches to address the complexities of human-machine interaction in the modern digital landscape.

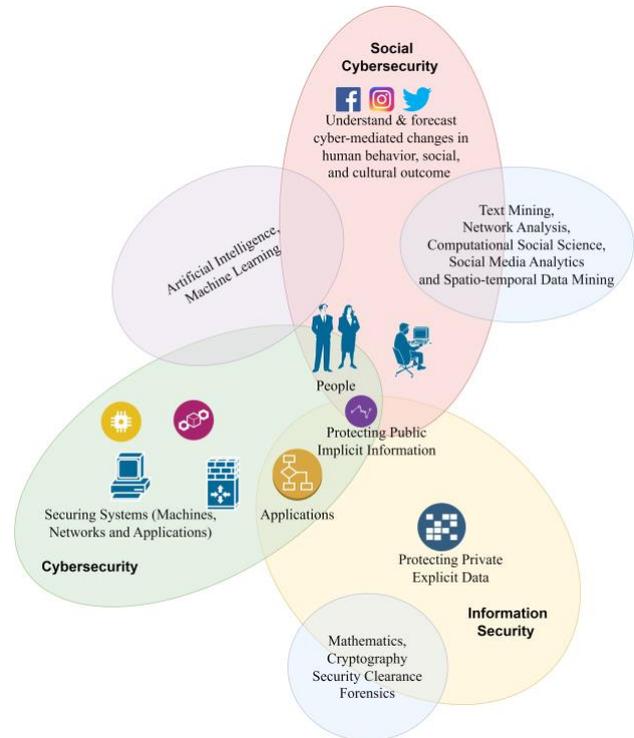

**Figure 1:** The relationship among cybersecurity, information security, and social cybersecurity.

Before venturing into predictions of social cybersecurity attacks, understanding the predictable elements and potential challenges is pivotal. The social dimension of cybersecurity, inherently complex due to human psychology and nuanced social interactions, presents significant forecasting challenges. Yet, given the shared tasks and use cases within this domain that exhibit similarities, it's possible to identify patterns. Such recognition of commonalities facilitates valuable insights and strategies to address these challenges.

While the primary goal is the detection of social cybersecurity threats, insights gained from predicting future attacks and forecasting methodologies offer valuable context. The unpredictable nature of human behaviors, group dynamics, and evolving societal norms [19; 20; 21; 22; 23] enriches our understanding of the threat landscape, thereby enhancing detection systems. This unpredictability underscores the necessity of incorporating a detailed grasp of attack statistics, the presence of threats, and the influence of social behaviors on these variables, even when our focus is on detecting, rather than forecasting, the security status of a social network.

Moreover, two pivotal components—*attack projection* [24], which aids in understanding the sequence of actions following an attack, and *intention recognition* [25], which involves discerning an adversary's intentions—can significantly inform detection strategies. Even though these components are traditionally associated with forecasting, they are equally critical in shaping robust detection mechanisms that can anticipate and respond to threats more effectively.

Despite the fluidity of social trends and interactions posing challenges, this survey emphasizes methodologies that not only advance forecasting but also contribute to the detection of cybersecurity threats. Many of these methodologies are grounded in shared theoretical foundations, offering a multidisciplinary approach to enhance detection capabilities in the face of social cybersecurity challenges.

### Key Contributions

This survey explores the detection of social cybersecurity attacks, providing valuable insights into the field. The main highlights of our work are:

- **Valuable Resource**: Serving as an essential guide for those venturing into social cybersecurity, this paper delves into diverse prediction methods. From machine learning to agent-based modeling, we provide a broad understanding of the field, especially in the context of the increasing threats in online networks.

- **Comprehensive Exploration**: Our survey covers a vast range of topics associated with social cybersecurity attacks. More than just identifying the challenges, we also suggest potential solutions, ensuring readers not only understand the problem but also the ways to address it.

- **Datasets and Tools**: A key aspect of research is the tools and data one employs. We highlight crucial public datasets and tools that researchers can leverage in the realm of social cybersecurity.

- **Research Challenges and Future Directions**: Research is not without its hurdles. We take a deep dive into the complexities and challenges faced in this domain, especially when dealing with social engineering threats. By shedding light on potential research avenues, we emphasize the ongoing need for innovation and exploration in the field.

By explicitly addressing the needs of researchers, policymakers, and industry professionals, our survey provides a comprehensive framework that not only advances academic knowledge but also informs decision-making in both regulatory and operational contexts. As the field of social cybersecurity continues to evolve, this work will serve as a foundation for interdisciplinary collaborations that integrate technological innovation, human factors, and governance strategies.

In essence, our paper furnishes a comprehensive overview of prediction methods, details specific threats, showcases pivotal datasets and tools, and outlines future research challenges and directions. This makes it an indispensable guide for those diving into the dynamic world of social cybersecurity.

### Literature Search Methodology and Paper Structure

In our initial quest for literature, we targeted journals with a penchant for survey-oriented articles on social cybersecurity predictions. Despite our focused approach, no papers exclusively dedicated to the forecasting and prediction of cyberattacks in this niche were identified. This observation led us to broaden our search horizon to include Google Scholar, IEEE Xplore, and the ACM Digital Library.

To ensure comprehensive coverage, we used a combination of search queries including: "social cybersecurity" OR "social cyber security" OR "Social-cybersecurity" OR "Cyber-social security" OR "SocCyber" OR "Social-network cybersecurity". These search terms were selected based on common terminologies used in the field and variations observed across existing literature. The search was conducted using title, abstract, and keyword-based filtering to identify relevant studies.

The selection criteria for papers included: (i) peer-reviewed articles published in reputable cybersecurity and social computing journals, (ii) studies focusing on prediction and detection methodologies within social cybersecurity, and (iii) papers providing novel insights into adversarial threats in digital environments. We excluded studies that focused solely on traditional cybersecurity threats without a social dimension or those lacking empirical validation. Additionally, we acknowledge the potential for selection bias, as research primarily indexed in the selected databases may not represent all relevant findings. However, we mitigated this risk by cross-referencing citations from key papers and reviewing recent conference proceedings in cybersecurity and social network analysis.

By doing so, we have endeavored to make our survey both comprehensive and reflective of the most recent and influential research on the forecasting and prediction of social cybersecurity attacks.

To assist in navigating the terminologies, Table 1 enumerates the crucial acronyms used throughout the paper.

The structure of the paper is outlined as follows: Section 2 provides an overview of the related work pertinent to our study. Section 3 presents a detailed analysis of various social cybersecurity attacks, along with their respective evaluations and potential solutions. Section 4 offers a deep dive into the taxonomy of techniques used for the detection of such attacks. The subsequent Sections, from 4.1 to 4.4, offer comprehensive literature reviews spanning Machine Learning (ML) methods, discrete models, metaheuristic algorithms, and agent-based modeling. Section 5 gives an analytical overview of the tools and publicly available datasets significant to social cybersecurity research. The paper reaches its denouement in Section 6, which synthesizes the current challenges and solutions in social cybersecurity and suggests avenues for future exploration. Finally, Section 7 provides the concluding remarks of our study.

**Table 1**
List of important acronyms.

| Acronym | Full Form | Acronym | Full Form |
|---|---|---|---|
| ACO | Ant Colony Optimization | LSH | Locality-sensitive Hashing |
| ALO | Ant Lion Optimization | LSTM | Long-Short Term Memory |
| AP | Affinity Propagation | LVSM | Linear Support Vector Machine |
| ARIMA | Autoregressive Integrated Moving Average | MC | Markov Chain |
| ARIMAX | RIMA with explanatory variables | MCC | Matthews Correlation Coefficient |
| BAT | Binary Bat Algorithm | MCC | Multi-Class Classification |
| BN | Bayesian Networks | MFO | Moth Flame Optimization |
| BOTRflCN | Bot detection with Relational flraph Convolutional Networks | ML | Machine Learning |
| BOW | Bag-of-Words | MLP | Multi-Layer Perception |
| BTO | Binary Term Occurrence | MLP | Multi-Layer Perception |
| CASOS | Computational Analysis of Social and Organizational Systems | MRF | Markov Random Field |
| CAT | CVE-Author-Tweet | MSIfl | Multi-Start Iterated flreedy |
| CDC | Centers for Disease Control and Prevention | MSRC | Microsoft Security Response Center |
| CNN | Convolutional Neural Network | NB | Naïve Bayes |
| CNN | Convolutional Neural Network | NDA | National Vulnerability Database |
| CSA | Cuckoo Search Algorithm | NER | Named Entity Recognizer |
| CUCKOO | Binary Cuckoo Algorithm | NLP | Natural Language Processing |
| CVE | Common Vulnerabilities and Exposures | NODEXL | Network overview, discovery, and exploration |
| CVSS | Common Vulnerability Scoring System | OSINT | Open-Source Intelligence |
| DDOS | Distributed Denial-of-Service | PCA | Principal Component Analysis |
| DNN | Deep Neural Network | PSO | Particle Swarm Optimization |
| DQE | Dynamic Query Expansion | QE | Query Expansion |
| DT | Decision Tree | RBF | Radial Basis Function |
| EM | Expectation-Maximization | RBF KERNEL | Radial Basis Function kernel |
| FA | Firefly Algorithm | RF | Random Forest |
| FEEU | Forecasting Ensemble for Exploit Timing | RflCN | Relational flraph Convolutional Network |
| FEEU-XflBOOST | FEEU with Extreme flradient Boosting (XflBoost) | RNN | Recurrent Neural Network |
| FRET | Forecasting Regression for Exploit Timing | SflD | Stochastic flradient Descent |
| flA | flenetic Algorithm | SMO | Sequential Minimal Optimization |
| flCN | flraph Convolutional Networks | SMOTE | Synthetic Minority Oversampling Technique |
| flDELT | fllobal Database of Events, Language, and Tone | SNA | Social Network Analysis |
| flMM | flaussian Mixture Models | SNAP | Stanford Network Analysis Project |
| flPU | flraphics Processing Unit | SOCNETV | Social Network Visualizer |
| flRU | flated Recurrent Unit classifier | SSO | Social Spider Optimization |
| HLT | Human Language Technologies | SVD | Singular Value Decomposition |
| HMM | Hidden Markov Model | SVM | Support Vector Machine |
| JUNfl | Java Universal Network/flraph | TDM | Text and Data Mining |
| KNN | K-nearest-neighbor | TF | Term Frequency |
| L-BFflS | Limited-memory Broyden-Fletcher-floldfarb-Shanno | T-MRF | Typed Markov Random Field |
| LBP | Local Binary Patterns | TO | Term Occurrences |
| LDA | Linear Discriminant Analysis | TSA | Tunicate Swarm Algorithm |
| LDA | Latent Dirichlet Allocation | UflM | Undirected flraphical Models |
| LDL | Latent Dirichlet Allocation | VERIS | Vocabulary for Event Recording and Incident Sharing |
| LFA | Levy flight Firefly Algorithm | WOA | Whale Optimization Algorithm |
| LIBSVM | Library for Support Vector Machines | WSA | Wolf Search Algorithm |
| LR | Logistic Regression | XflBOOST | Extreme flradient Boosting |

To provide clarity and focus to this survey, we aim to explore the general landscape of detecting social cybersecurity attacks, emphasizing methodologies that identify threats arising from manipulative digital activities. The scope of this survey is framed by the following central research questions:

- What are the primary challenges in detecting social cybersecurity threats?

- What tools and techniques are most effective in addressing these challenges?

- How can insights from forecasting methods enhance detection capabilities?

By clearly delineating these boundaries, this study focuses on identifying practical and theoretical approaches to improve detection mechanisms within the complex landscape of social cybersecurity.

## 2. Related Works

Our research primarily addresses the general landscape of detecting social cybersecurity attacks. We uniquely fill this gap, laying a foundation for proactive defense strategies. Several surveys on social cybersecurity exist, but they don't concentrate on the detection in the depth that we do. A distinction is illustrated in Table 2, utilizing abbreviations like ML (Machine Learning), DM (Discrete Models), MH (Metaheuristic Algorithms), and ABM (Agent-Based Modeling).

One group of papers centered on social network analysis brings to light several findings. Kirichenko et al. [26] delve into social network techniques for cyber threat detection, yet their scope seems narrow. Pavel et al. [27], even with a broader cybersecurity perspective, frequently neglect social attacks. [28] offers a profound look into cybersecurity within social networks but does not accentuate forecasting or prediction of attacks. Carley's work [5], although vast in its coverage of social cybersecurity, misses the mark on the prediction front.

**Table 2**
Comparison with existing surveys.
(**Legends**. ML: Machine Learning; DM: Discrete Models; MH: Metaheuristic Algorithms; ABM: Agent-Based Modeling.)

| Ref. | Year | Social Cybersecurity Attack Prediction and Forecasting Techniques | | | | Attacks, Evaluation, and Solutions | Challenges | Future Directions | Dataset | Tools | Overview |
|---|---|---|---|---|---|---|---|---|---|---|---|
| | | ML | DM | MH | ABM | | | | | | |
| [26] | 2018 | ✓ | ✓ | ✗ | ✗ | ✗ | ✗ | ✗ | ✗ | ✗ | This survey paper explores the detection of cyber threats through social networks, yet it offers limited coverage and misses opportunities to provide comprehensive insights |
| [27] | 2019 | ✓ | ✓ | ✗ | ✗ | ✗ | ✓ | ✗ | ✗ | ✗ | The survey paper underscore projection, prediction, and forecasting in cybersecurity, yet they often neglect social attacks and lack a thorough grasp of social engineering |
| [28] | 2019 | ✗ | ✗ | ✗ | ✗ | ✗ | ✗ | ✗ | ✗ | ✗ | This survey paper explores improvements in cybersecurity and privacy within social networks, but it doesn't delve into the prediction or forecasting of social cybersecurity attacks |
| [5] | 2020 | ✗ | ✗ | ✗ | ✗ | ✗ | ✗ | ✓ | ✗ | ✗ | This survey paper gives an overview of social cybersecurity research, yet it does not adequately focus on predicting and forecasting social cybersecurity attacks |
| [29] | 2020 | ✓ | ✗ | ✗ | ✗ | ✗ | ✓ | ✓ | ✗ | ✗ | This survey paper delves into ML techniques for cybersecurity but doesn't address the prediction and forecasting of social cybersecurity attacks, also overlooking the aspects of social engineering |
| [30] | 2022 | ✓ | ✗ | ✗ | ✗ | ✗ | ✓ | ✓ | ✓ | ✗ | This survey paper investigates the application of ML in cybersecurity, with a significant emphasis on data and malware detection, but it doesn't adequately address the prediction and forecasting of social cybersecurity attacks |
| [31] | 2021 | ✓ | ✗ | ✗ | ✗ | ✗ | ✓ | ✓ | ✓ | ✗ | This survey paper explores spam detection on social networks using various methods, offering detailed insights into spam detection techniques. However, it has a narrow focus on spam detection and does not specifically address the prediction and forecasting of social cybersecurity attacks |
| [32] | 2015 | ✓ | ✗ | ✗ | ✗ | ✗ | ✓ | ✓ | ✓ | ✗ | This survey paper reviews ML and DM methods for cyber analytics in intrusion detection, providing tutorial-like descriptions. It addresses challenges in applying ML/DM for cyber security but lacks exploration of prediction and forecasting in social cybersecurity attacks and methods, leaving room for further research |
| [33] | 2023 | ✓ | ✗ | ✗ | ✗ | ✗ | ✓ | ✓ | ✓ | ✗ | This survey paper investigates smart grid vulnerabilities and suggests ML and blockchain-based security solutions. While it effectively addresses cyberattack challenges, it lacks exploration of prediction and forecasting of social cybersecurity attacks, calling for further research in understanding ML's role in anticipating such threats |
| [34] | 2023 | ✓ | ✗ | ✗ | ✗ | ✗ | ✓ | ✓ | ✓ | ✗ | This survey paper explores unsupervised learning methods for cyber detection in smart grids, specifically focusing on identifying False Data Injection Cyber Attacks (FDIA). However, it does not cover prediction and forecasting methods for social cybersecurity attacks, leaving room for further research in understanding the role of different learning approaches in anticipating such threats |
| Our survey paper | 2025 | ✓ | ✓ | ✓ | ✓ | ✓ | ✓ | ✓ | ✓ | ✓ | Our survey explores the detection of social cybersecurity threats, interweaving a variety of techniques and attack vectors with their respective countermeasures. We provide an in-depth examination of tools and datasets, address current challenges, suggest potential solutions, and advocate for continued research to advance this critical domain. |

A group focusing on Machine Learning (ML) in cybersecurity presents varied insights. Kamran et al. [29] traverse the expanse of ML techniques across different cybersecurity domains, ranging from spam detection to 5G security, but bypass the prediction of social cybersecurity attacks. On a similar trajectory, Dasgupta et al. [30] delve deep into ML's applications in cybersecurity yet omit crucial details on forecasting and prediction of social threats.

Turning to spam and intrusion detection, another group offers its perspective. Rao et al. [31] provide a meticulous survey on spam detection strategies in social networks, but their lens remains predominantly on spam. Buczak et al. [32] offers an exhaustive review of ML and data mining techniques tailored for intrusion detection but skips the intricacies of social cybersecurity prediction and forecasting.

In the realm of smart grids, a distinct group emerges. Mazhar et al. [33] shed light on vulnerabilities and propose ML and blockchain-driven security solutions but overlook prediction methods for social threats. Pinto et al. [34] navigate the domain of cybersecurity in smart distribution systems, predominantly leveraging unsupervised learning methods but disregarding the prediction and forecasting of social attacks.

While previous surveys have made valuable contributions to the study of social cybersecurity, they exhibit notable gaps that necessitate further research. As seen in Table 2, many surveys provide broad overviews of cybersecurity without focusing on prediction and forecasting techniques specific to social cybersecurity attacks. For example,

Kirichenko et al. [26] explore social network-based cyber threat detection but lack predictive modeling approaches that could proactively identify evolving threats. Similarly, Pavel et al. [27] discuss forecasting in cybersecurity but do not extend this discussion to social engineering-based attacks, which have distinct behavioral patterns and spread mechanisms.

Existing ML-based cybersecurity surveys, such as Kamran et al. [29] and Dasgupta et al. [30], predominantly focus on traditional intrusion detection and malware analysis but do not integrate forecasting models to anticipate adversarial strategies in social cybersecurity. Even comprehensive cybersecurity reviews such as those by Buczak et al. [32] and Verma et al. [31] emphasize spam and intrusion detection while overlooking forecasting and real-time adaptation of attack prediction models.

Furthermore, limited emphasis has been placed on the role of interdisciplinary approaches, such as social cyberforensics and dynamic network analysis, which can enhance predictive capabilities in social cybersecurity. Notably, prior surveys do not explore how hybrid techniques—combining ML, agent-based modeling (ABM), and metaheuristic algorithms—can be leveraged for proactive threat identification. Additionally, a lack of publicly available datasets and benchmarking frameworks has been identified as a constraint in developing reliable detection systems, a challenge that our work addresses.

By addressing these gaps, our survey provides a comprehensive taxonomy of detection techniques, spanning ML, discrete models, metaheuristic optimization, and ABM, while also identifying future directions in predictive analytics for social cybersecurity. We uniquely focus on the integration of real-time threat prediction and countermeasure optimization, distinguishing our work from prior studies.

In summing up our contribution, we shine a light on the detection strategies for social cybersecurity attacks. We weave a cohesive narrative that connects a spectrum of techniques and attack vectors with their corresponding countermeasures, along with an analysis of tools and datasets. Furthermore, we unveil the existing challenges within the realm of social cybersecurity, offering contemporary solutions and proposing directions for future inquiry. Significantly, our survey identifies uncharted territories such as social cyberforensics and dynamic network analysis, signposting these as promising fields for forthcoming research endeavors.

## 3. Various Attacks on Social Cybersecurity: Evaluations and Potential Solutions

In this survey paper, we deeply explore various attacks observed within the realm of social cybersecurity, each posing substantial risks to user privacy and security, profoundly affecting individuals and communities.

Exploration delves deeper than just outlining the nature and methodologies of these attacks. We provide succinct evaluations, dissecting each attack's implications, challenges, and potential repercussions. These evaluations shed light on the multifaceted risks and their impacts on societal harmony, individual well-being, and overall cybersecurity. This provides an insightful understanding of the intricate challenges posed by social cybersecurity threats.

Furthermore, we present potential solutions aimed at reducing the occurrence and impact of the highlighted attacks. These solutions, encompassing preventative measures, remedial strategies, and insightful recommendations, address the core of the identified issues. They offer strategies to counter the widespread malicious activities in social cybersecurity. Central to these solutions is the idea of fostering a secure and inclusive online environment. This is achieved through enhanced user awareness, strong policy frameworks, and shared responsibility.

For a quick reference and a comprehensive overview of the discussed attacks, evaluations, and proposed solutions, readers are directed to Table 3. This compilation serves as an invaluable resource, aiding in understanding the challenges introduced by social cybersecurity attacks and suggesting effective counter-strategies. Also, Figure 2 shows the mechanism of social cyber attack detection and evaluation.

**Identity Theft**:

Identity Theft [35] occurs when an unauthorized entity illegitimately gains access to a user's account and exploits it, potentially causing harm to the victim through theft of sensitive information or dissemination of malicious content. This form of attack is particularly insidious as it allows the attacker to operate under the guise of a legitimate user, thus enhancing the credibility of their malicious activities.
**Evaluation:** [36] The degree of harm and the scope of exploitation in identity theft are considerable, given that the attacker can misuse the victim's account to access sensitive data or spread malevolent content. The clandestine nature of this attack makes it a potent threat, with the victims often remaining oblivious until significant damage has occurred, highlighting the critical need for early detection and preventive measures.
**Potential Solutions:** [37] Preventing identity theft necessitates the implementation of robust authentication methods and continuous user education on secure password practices and the importance of maintaining account security. Employing multi-factor authentication and regularly updating passwords can significantly reduce the risk of identity theft. Additionally, creating awareness about the various tactics employed by attackers can equip users with the knowledge to identify and avoid falling prey to such attacks, thereby enhancing overall security in online social interactions and networks.

**Spam Attack**:

Spam Attacks [38] involve attackers acquiring a user's contact information to send unsolicited spam emails, leading to network congestion and causing potential inconveniences to service providers and users. The indiscriminate nature of these attacks can overwhelm users with irrelevant or malicious content, disrupting the normal flow of communications on the platform.

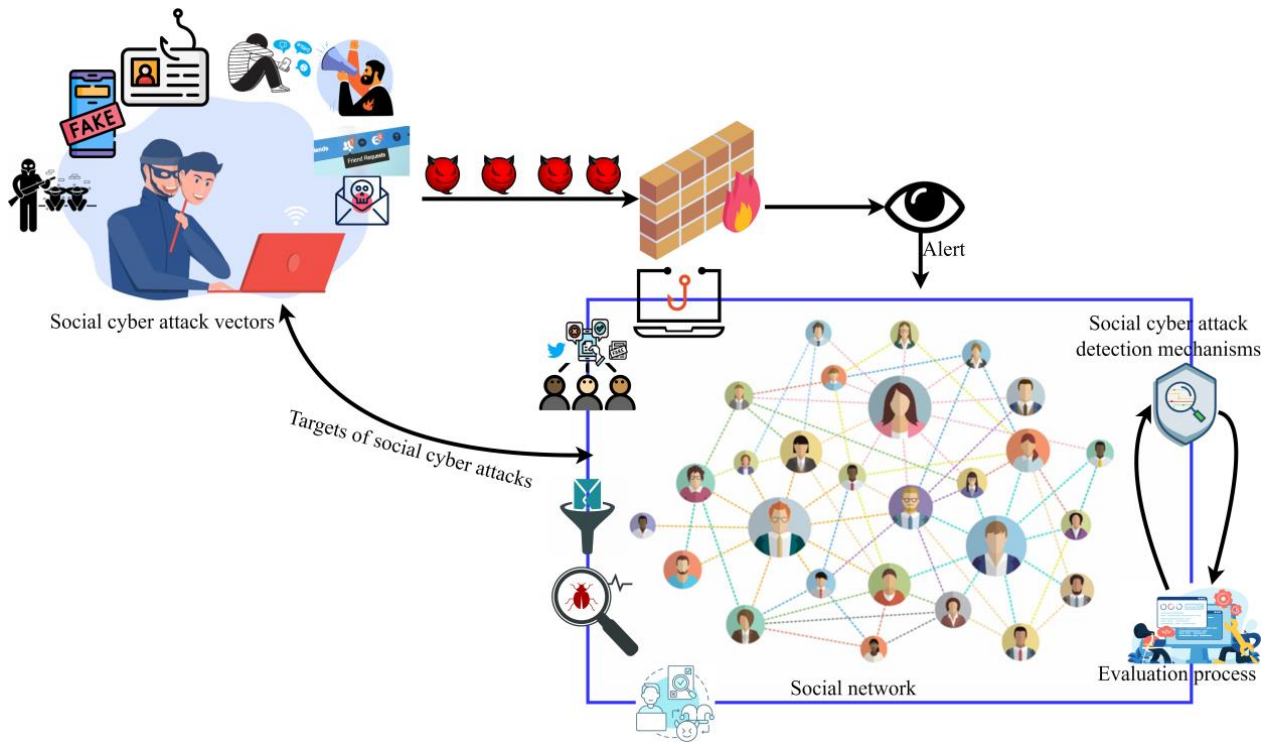

**Figure 2:** The Mechanism of social cyber attack detection and evaluation.

**Evaluation:** [39] The ubiquity of spam attacks can adversely affect the user experience on online platforms by cluttering feeds with unwanted content, potentially overshadowing legitimate communications and impairing the platform's functionality. The malicious intent behind some spam attacks further intensifies the risks associated with them, with potential damages including exposure to harmful content or websites and loss of personal information.

**Potential Solutions:** [40] In the context of social cybersecurity, mitigating spam attacks is pivotal. This can be achieved through the implementation of advanced email filters and anti-spam measures, ensuring effective blocking of unwanted content. Encouraging users to exercise caution, especially when confronted with unexpected communications, can aid in creating a vigilant online community. Equipping them with the requisite knowledge to identify and promptly report spam further strengthens the defense against such attacks. Moreover, the continuous refinement and enhancement of spam detection algorithms play a crucial role in preserving the integrity and ensuring the user-friendly nature of online platforms and communities.

**Malware Attack:**

Malware attacks [41] involve the distribution of harmful scripts or software via networking sites, which can result in the unauthorized installation of malware on a user's device or lead to the theft of personal data. These attacks exploit the interconnectedness and trust within platforms to propagate malicious software, compromising user privacy and security.

**Evaluation:** [42] The interconnected nature of online platforms and communities in the realm of social cybersecurity makes them susceptible to the rapid spread of malware, resulting in significant harm to users and the platform's infrastructure. The potential consequences of such attacks encompass the loss of sensitive data, unauthorized system intrusions, and eroded user trust in these digital communities and platforms.

**Potential Solutions:** [43] To counter malware attacks in online communities and platforms, it's crucial to educate users about safe browsing habits and the dangers of downloading content from unverified sources. Implementing and regularly updating anti-malware software can assist in detecting and neutralizing malicious content before any damage occurs. Further, fortifying security measures and integrating advanced threat detection mechanisms within these digital platforms and networks can facilitate early recognition and response to malware threats, ensuring both user data protection and sustained platform integrity.

**Sybil Attack:**

Sybil Attacks [44] refer to the creation and use of fake identities or profiles within online communities and platforms. Attackers utilize these counterfeit profiles to spread misinformation, disseminate malware, or disrupt the functionality of the platform. Such attacks leverage the built-in trust within digital networks, enabling the malicious entity to manipulate or deceive unwary users.

**Evaluation:** [45] The inherent nature of online communities and platforms, built on principles of openness and connectivity, renders them especially susceptible to Sybil Attacks. An individual attacker can manage numerous fabricated profiles, amplifying the potential damage. Such attacks not only interrupt genuine communication pathways but also undermine the perceived reliability and authenticity of information within these digital spaces.

**Potential Solutions:** [46] Effective countermeasures against Sybil Attacks involve improving user registration and authentication procedures. Implementing more stringent registration protocols, like CAPTCHAs, phone number verification, or behavior-based checks, can act as deterrents for automated account creation. Regular monitoring and algorithmic detection of suspicious activity patterns can also aid in identifying and removing fake profiles. Raising user awareness about the prevalence of fake accounts and encouraging them to report suspicious behavior can further bolster platform defenses against Sybil Attacks.

**Exploiting Community Detection:**

Community detection has many applications in online networks [47]. Exploiting community detection [48] involves malevolent activities aimed at the intrinsic structure and communication dynamics of these digital communities. Through such attacks, adversaries can identify and target susceptible groups or communities within online platforms. For example, a digital group of users discussing sensitive health information might be singled out by malicious actors aiming to penetrate and misuse the group's collective resources or data.

**Evaluation:** [49] The capability to cluster and identify communities within online platforms might unintentionally make these groups vulnerable to specialized attacks. Certain communities could be more attractive to attackers because of the type of content they share, the potential for data extraction, or chances to circulate false information. Such attacks don't just jeopardize the privacy and safety of community participants; they can also diminish trust within the community, rendering sincere exchanges permeated with doubt.

**Potential Solutions:** [50] Defending against the exploitation of community detection requires enhancing the security protocols governing group interactions and membership. This might involve more rigorous membership vetting processes, tighter content sharing restrictions, and regular monitoring for suspicious activities. Additionally, advanced algorithms can be developed to identify and flag potential infiltrators or malicious actors. Educating community moderators about potential threats and equipping them with tools to monitor and remove harmful entities is equally crucial. Such proactive and reactive measures can significantly bolster community defenses and preserve the integrity and purpose of these online groups.

**Social Phishing:**

Social Phishing [51] is a deceptive technique where attackers create fraudulent websites or communication channels that appear legitimate, aiming to deceive users into providing personal information or login credentials. These attacks leverage the trust that users place in familiar platforms or known contacts, effectively manipulating them into compromising their own security.

**Evaluation:** [52] The inherent danger of social phishing stems from its ability to exploit the trust users have in familiar entities. Because these phishing attempts often masquerade as trusted sources or acquaintances, they can easily bypass users' typical guard against suspicious activities. The repercussions of such attacks can be severe, with users potentially losing personal data and financial resources or facing identity theft.

**Potential Solutions:** [53] Combatting social phishing necessitates a two-pronged approach: technological interventions and user education. On the technological front, implementing robust phishing detection algorithms and secure communication protocols can identify and block suspicious content or redirect attempts. Moreover, periodic system-wide security audits can ensure that vulnerabilities are promptly addressed. On the user front, continuous education about the dangers of phishing and training on how to recognize phishing attempts are vital. Encouraging users to verify suspicious messages through alternate communication channels and to avoid clicking on unverified links can significantly reduce the risk of successful phishing attacks.

**Impersonation:**

Impersonation [54] denotes the act where an attacker establishes a counterfeit profile within online platforms or communities, emulating a legitimate individual or organization. The primary motive behind these attacks is to mislead other participants, utilizing the trust linked to the mimicked identity to disseminate false information, gather private data, or conduct various harmful actions.

**Evaluation:** [55] Impersonation attacks capitalize on the intrinsic trust and recognition dynamics within online communities and platforms. Users are often more forthcoming when engaging with recognized entities, rendering them vulnerable to deception when encountering a forged account. Such transgressions can result in numerous negative outcomes, such as the circulation of erroneous information, invasions of privacy, and possible financial deceit.

**Potential Solutions:** [56] Countering impersonation requires a combination of robust platform policies, technological solutions, and user awareness. Platforms can implement more stringent verification processes for accounts, especially those with significant reach or influence. Features such as "Verified Badges" can help users distinguish genuine profiles from potential impersonators. Additionally, utilizing advanced ML algorithms to detect and flag potential impersonation activities based on account behavior and content can be instrumental. On the user side, education campaigns highlighting the risks of impersonation and guiding users on how to verify account authenticity can help reduce the success rate of such attacks.

**Hijacking:**

Hijacking [57] entails an attacker forcefully seizing control of a user's authentic online profile or account, generally by cracking or circumventing their access credentials. Upon gaining control, the attacker can exploit the account for diverse nefarious activities, from spreading misleading data to impersonating the genuine user for deceitful objectives.

**Evaluation:** [58; 59] Account hijacking poses significant threats to both the direct victim and their network of contacts. The credibility and trust associated with a hijacked account can amplify the impact of malicious actions taken by the attacker. Additionally, the direct victim might face loss of personal data, privacy breaches, and potential reputational harm, especially if the attacker engages in activities that misrepresent the victim's intentions or beliefs.

**Potential Solutions:** [60] Preventing account hijacking necessitates the adoption of robust authentication methods. Incorporating multi-factor authentication, where users are required to provide two or more verification methods, can significantly bolster account security. Platforms can also implement mechanisms to detect unusual account activities, such as logging in from new locations or devices, prompting immediate verification processes in such cases. Educating users about the importance of strong, unique passwords and the risks of using the same password across multiple platforms can also reduce the likelihood of successful hijacking attempts.

**Fake Requests**:

Fake Requests [61] involve attackers using counterfeit profiles to send deceptive connections or information requests to other users. By expanding their network through these fake requests, attackers can gain access to a wider range of personal data, increasing their potential for malicious activities or data harvesting.

**Evaluation:** [62] The core danger of fake requests is the potential breach of privacy. Unsuspecting users might share sensitive information, either directly or indirectly, with these fake accounts, making them vulnerable to subsequent attacks or misuse of their data. Additionally, accepting such requests can lead to exposure to misleading content, spam, or even direct phishing attempts.

**Potential Solutions:** [63] Combating fake requests necessitates both platform-level interventions and heightened user awareness. On the platform side, implementing more rigorous profile verification procedures and using algorithms to detect and flag potential fake accounts can play a crucial role in minimizing the spread of such requests. For users, it's essential to be educated on the risks of accepting unknown requests and to be encouraged to adopt a cautious approach, only accepting connections from familiar or verified entities. Additionally, platforms can provide users with tools to easily report suspicious requests, facilitating faster identification and removal of fake profiles.

**Image Retrieval and Analysis**:

This type of attack [64] leverages cutting-edge face and image recognition technologies. Malevolent actors can extract and scrutinize images from online profiles or communities to acquire insights about an individual, their pursuits, affiliations, and even regular locations. By assembling this data, attackers can infringe upon the privacy and safety of the individual, extending the risk to their associated contacts, acquaintances, and relatives.

**Evaluation:** [65] Image Retrieval and Analysis present multifaceted dangers. Beyond the straightforward breach of privacy, the depth of extracted data can fuel targeted phishing schemes, identity fraud, or even real-world tracking. Considering the prevalent dissemination of images in online communities and platforms, numerous participants might be oblivious to the magnitude of information that can be deduced and studied from seemingly harmless snapshots.

**Potential Solutions:** [66] Mitigating the risks associated with image retrieval and analysis demands a combination of technological solutions and user awareness initiatives. Platforms can provide enhanced privacy settings, allowing users to control who can view and access their images, along with implementing automated filters to deter image scraping attempts. On the user end, awareness campaigns detailing the risks of sharing personal images and the potential information they can reveal can encourage more judicious sharing practices. Additionally, leveraging technologies like image obfuscation or watermarking can deter unauthorized analysis or replication of user photos.

**Cyberbullying**:

Cyberbullying [67] involves utilizing online communities and platforms to intimidate, threaten, belittle, or single out another individual. It includes sharing disparaging images, texts, or remarks with the intent to inflict emotional harm on the target. The electronic essence of these aggressions enables swift propagation, rendering them particularly damaging and widespread.

**Evaluation:** [68; 69] The impact of cyberbullying is profound, affecting the mental well-being of victims, leading to feelings of isolation, depression, and in extreme cases, even suicidal tendencies. The anonymity that online platforms offer can embolden bullies, making it challenging to identify and penalize them. Moreover, the vast reach and permanence of online content mean that bullying incidents can have lasting and far-reaching consequences.

**Potential Solutions:** [70] Tackling cyberbullying necessitates an all-encompassing strategy, merging platform-level initiatives, legislative actions, and collective community endeavors. Online platforms can incorporate rigorous content moderation utilities, facilitating the prompt detection and deletion of harmful content. Instituting procedures for users to report bullying events and offering support to the impacted parties is pivotal. Legal infrastructures can be set up to sanction cyberbullying, serving as a preventive measure. Moreover, enlightening the public about the detrimental consequences of cyberbullying and cultivating an ethos of understanding and regard in digital spaces can markedly mitigate such occurrences.

**Hate Speech**:

Hate speech [71] is defined by messages that denigrate, menace, or belittle individuals or collectives, typically zeroing in on aspects like race, faith, ethnicity, sexual inclination, disability, or gender. Its widespread presence in online communities and platforms is notably alarming given its vast influence and the profound unease it can elicit. Interpreting hate speech in the diverse landscape of online spaces can be complex, as cultural and societal subtleties influence linguistic perspectives. Yet, the intrinsically degrading and disparaging essence of hate speech earmarks it as a malicious form of communication.

**Evaluation:** [72] The vast influence of online communities can magnify the detrimental repercussions of hate speech, disseminating unease and potentially kindling hostility or bias. The intricacies in delineating hate speech, accentuated by the instantaneous, fluid exchanges in online platforms, present considerable hurdles in pinpointing and supervising such content. This highlights the urgency for holistic strategies to manage it aptly.

**Potential Solutions:** [73; 74]To mitigate the prevalence of hate speech, the development and implementation of robust content moderation policies and technologies, including AI-driven tools, are crucial. These tools can aid in the timely identification and removal of hate speech. Additionally, fostering an environment that encourages users to report hate speech and providing clear guidelines on acceptable content can contribute to creating a more inclusive and respectful online community. Educative initiatives aimed at promoting tolerance, diversity, and respect can also play a significant role in combating the propagation of hate-filled narratives and fostering a sense of community and shared responsibility among users.

### Terrorist Activity:

Terrorist Activity in online communities [75] pertains to the exploitation of these digital spaces by extremist factions for radicalization, enlistment, and the spread of their doctrines. Numerous extremist assemblies have discerned the potency of platforms like video-sharing sites, social networking sites, and microblogging services in accessing an international viewership and consequently manipulating them to advance their objectives.

**Evaluation:** [76; 77] The harnessing of online platforms by extremist entities presents grave risks to worldwide safety. Through these digital spaces, extremist views can disseminate swiftly, zeroing in on vulnerable individuals for enlistment and radicalization. The instantaneous and interlinked nature of online communities renders them a formidable instrument for these factions to orchestrate endeavors, broadcast propaganda, and even strategize assaults, potentially culminating in tangible aggression and turmoil.

**Potential Solutions:** [78] Addressing terrorist undertakings in online platforms necessitates cooperative endeavors from digital service providers, governmental entities, and global organizations. Platforms can bolster their content moderation strategies and employ AI-driven mechanisms to promptly discern and eliminate extremist material. Exchanging intelligence between platforms and with policing agencies can further assist in pinpointing and quelling threats. State authorities and international consortia can draft regulatory edicts mandating platforms to counteract extremist content and concurrently champion counter-narrative initiatives to contest extremist dogmas. Additionally, enlightening users about indications of radicalization and championing digital discernment can act as prophylactic strategies.

### Social Unrest:

Social Unrest in online communities [79] embodies coordinated collective endeavors that question prevailing standards and may disturb societal or institutional routines. These endeavors can take the form of public disruptions, widespread demonstrations, or even workforce protests. Displeasure with extant social, economic, or political circumstances can incite such movements, all aiming to champion transformation.

**Evaluation:** [80] The digital domain has revolutionized the way social unrest is orchestrated and disseminated. While online communities offer a formidable avenue for articulating grievances and rallying collective endeavors, they also introduce complexities. The swift proliferation of data, encompassing both authentic and spurious content, can heighten emotions, occasionally escalating serene demonstrations into tumultuous altercations. The distributed character of these communities complicates the verification of information integrity, ushering in potential disinformation and exploitation.

**Potential Solutions:** [81] To tackle the challenges posed by social unrest in online communities, platforms can institute protocols to validate the authenticity and consistency of disseminated information. Employing fact-checking utilities and affiliations with autonomous verification bodies can prove pivotal in curtailing disinformation. Moreover, nurturing transparent conversations and carving pathways for non-violent articulation of concerns can assist in addressing the foundational triggers of unrest. Platforms might also ally with civic entities to champion digital literacy, assuring users are adept at distinguishing trustworthy inputs from potentially deceptive narratives.

### Attack Ad:

An Attack Ad [82] is a form of political advertisement primarily designed to discredit or malign an opposing candidate or party. Leveraging various digital channels within the realm of social cybersecurity, these ads highlight perceived flaws, weaknesses, or controversial stances of the opponent, intending to shift public opinion against them. Often, within these cyber environments, ads might employ selective data, exaggeration, or even misinformation to achieve their objectives.

**Evaluation:** [83] The pervasive nature of digital channels in social cybersecurity amplifies the reach and impact of attack ads, allowing them to swiftly influence vast cyber communities. While political discourse and critique are inherent to democratic processes, the malicious or deceptive nature of some attack ads can distort public perception, undermine trust in democratic institutions, and polarize digital communities. The real-time spread and potential virality of such content within cyber environments further complicate efforts to ensure fair and truthful political campaigning.

**Potential Solutions:** [84] To mitigate the negative impacts of attack ads, platforms can implement stringent ad-review policies, ensuring that political advertisements meet specific standards of truthfulness and fairness. Collaboration with independent fact-checkers can help in verifying the authenticity of claims made in such ads. Additionally, transparency initiatives, such as ad libraries detailing the sponsors, reach, and target audience of political ads, can provide users with context and promote informed decision-making. Public awareness campaigns educating users about the nature of attack ads and encouraging critical media consumption can also play a pivotal role in fostering a more informed electorate.

**Fake News**:

Fake News [85] pertains to the deliberate spread of false or misleading information crafted to deceive and manipulate readers or viewers. Leveraging various media platforms, especially social media, fake news aims to distort reality, influence public opinion, sow discord, or advance specific agendas. Whether politically, economically, or socially motivated, fake news exploits the rapid dissemination potential of digital platforms to reach and mislead vast audiences.

**Evaluation:** [86] The proliferation of fake news on social media is particularly concerning due to the platform's real-time, widespread reach. Misinformation can quickly gain traction, creating a cascade of reinforced false beliefs among users. This not only erodes trust in credible information sources but also threatens democratic processes, public safety, and social harmony. The decentralized and user-generated nature of content on social media complicates efforts to discern and combat fake news effectively.

**Potential Solutions:** [87] Addressing the challenge of fake news requires a multi-faceted approach. Platforms can incorporate advanced AI-driven fact-checking tools to identify and flag potentially misleading content. Collaborations with independent fact-checking organizations can enhance the credibility of these efforts. Transparency features, like source verification badges or context panels, can provide users with additional information to gauge the reliability of content. On the user front, digital literacy campaigns that equip users to critically evaluate online content can be instrumental in mitigating the spread and impact of fake news. Furthermore, promoting a culture of cross-referencing information and relying on multiple trusted sources can help in fostering an informed and discerning online community.

**Deepfake Manipulation**:

Deepfake manipulation refers to the use of AI-generated synthetic media, typically created using deep learning techniques such as Generative Adversarial Networks (GANs), to fabricate audio, video, or images that appear authentic [88]. Deepfakes have been widely used in misinformation campaigns, political manipulation, and character defamation. Attackers leverage these techniques to distort reality, influence public perception, and damage reputations.

**Evaluation:** [89] Deepfakes pose a significant threat to online communities due to their realism and rapid dissemination through social media platforms. Their usage can result in the spread of false narratives, erosion of public trust, and manipulation of public opinion. The difficulty in distinguishing between real and manipulated content exacerbates the problem, creating fertile ground for misinformation and disinformation campaigns.

**Potential Solutions:** [90] Combating deepfakes requires the development of advanced detection algorithms capable of identifying synthetic media artifacts. Techniques such as facial landmark detection, frequency domain analysis, and deep learning-based classifiers have shown promise. Moreover, promoting media literacy and awareness among users is critical in reducing the impact of deepfakes. Collaborative efforts between platforms, researchers, and policymakers can further strengthen the ecosystem against deepfake threats.

**AI-Generated Misinformation**:

AI-generated misinformation involves the use of large language models (LLMs) and content generation algorithms to create and propagate false or misleading information [91]. Unlike traditional misinformation, AI-generated content can be rapidly produced, highly convincing, and tailored to specific audiences, making it a powerful tool for adversaries in social cybersecurity contexts.

**Evaluation:** [92] The proliferation of AI-generated misinformation has significantly raised concerns due to its scale and quality. Attackers can automate the creation of fake news articles, social media posts, and bot-driven interactions, contributing to information overload and undermining trust in legitimate information sources. The capacity of AI models to mimic human-like communication further amplifies the threat.

**Potential Solutions:** [87] Addressing AI-generated misinformation requires a multi-faceted approach combining AI-based detection methods, fact-checking systems, and platform-level interventions. Detection models can be trained to recognize linguistic patterns, content anomalies, and metadata inconsistencies typical of generated misinformation. Moreover, strengthening user education and promoting fact-checking practices are essential in limiting the influence of such misinformation.

## 4. Techniques for Detection in Social Cybersecurity

This section delineates a taxonomy of methods employed in detecting social cybersecurity attacks. These methods span a broad spectrum, varying based on both their application domains and their mathematical and theoretical underpinnings. Previous surveys in the field of social cybersecurity [28; 5] have primarily revolved around enhancing cybersecurity and privacy in social networks, offering a broad overview of social cybersecurity research. Our survey, however, brings a unique focus on detecting social cybersecurity attacks, categorizing these methods based on their theoretical foundations. This approach underscores the commonalities across different tasks. Each method is additionally examined in terms of its specific use case, offering a nuanced understanding. The techniques for attack detection methods that emerge from this categorization are visually represented in Figure 3.

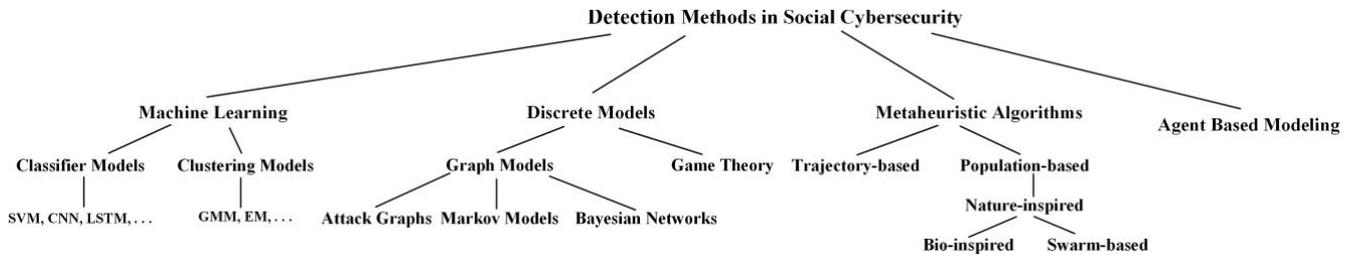

**Figure 3:** Techniques for detection methods in social cybersecurity.

**Table 3**
Summary of social media attacks: evaluation and potential solutions

| Attack | Evaluation | Potential Solution |
|---|---|---|
| Identity Theft [35] | Considerable harm due to misuse of victim's account, often with victims unaware until significant damage occurs [36] | Robust authentication, user education, multi-factor authentication, and regular password updates [37] |
| Spam Attack [38] | Clutters user feeds with unwanted content, impairing platform functionality and exposing users to potential threats [39] | Advanced email filters, anti-spam measures, and user vigilance [40] |
| Malware Attack [41] | Rapid malware dissemination compromises user privacy and security, leading to data theft and reduced trust [42] | User education, up-to-date anti-malware software, enhanced security protocols, and threat detection [43] |
| Sybil Attack [44] | Amplified potential harm due to operation of multiple fake profiles, undermining platform trustworthiness [45] | Stringent registration protocols, continuous monitoring, and user education [46] |
| Exploiting Community Detection [47; 48] | Targeted attacks compromise group privacy and erode trust within communities [49] | Enhanced security protocols for groups, rigorous membership vetting, and content restriction [50] |
| Social Phishing [51] | Exploits user trust in familiar platforms, leading to data loss and identity theft [52] | Robust phishing detection, system-wide security audits, and continuous user education [53] |
| Impersonation [54] | Uses trust dynamics to spread misinformation or solicit data, leading to potential financial fraud [55] | Stringent profile verification procedures, ML algorithms, and user education [56] |
| Hijacking [57] | Compromises direct victims and their network, leading to potential data theft and reputational harm [58; 59] | Robust authentication, multi-factor authentication, user education on strong passwords, and detection of unusual activities [60] |
| Fake Requests [61] | Privacy breaches due to malicious connections, leading to exposure to misleading content [62] | Rigorous profile verification, detection algorithms, user education, and reporting tools [63] |
| Image Retrieval and Analysis [64] | Invasion of privacy and potential for targeted attacks, identity theft, or physical stalking [65] | Enhanced privacy settings, image obfuscation, watermarking, and user awareness campaigns [66] |
| Cyberbullying [67] | Severe impact on mental well-being, isolation, depression, and potentially suicidal tendencies [68; 69] | Stringent content moderation, support systems for victims, legal penalties, and community awareness campaigns [70] |
| Hate Speech [71] | Amplifies harmful impacts, potentially inciting violence or discrimination, with challenges in content moderation [72] | Robust content moderation, AI-driven tools, user reporting, and educative initiatives [73; 74] |
| Terrorist Activity [75] | Threatens global security by spreading radical ideologies, targeting individuals for recruitment, and potential real-world violence [76; 77] | Enhanced content moderation, intelligence sharing, legal frameworks, and counter-narrative campaigns [78] |
| Social Unrest [79] | Amplifies sentiments, potentially escalating peaceful protests into violence, with misinformation risks [80] | Fact-checking tools, open dialogues, digital literacy campaigns, and partnerships with civic organizations [81] |
| Attack Ad [82] | Distorts public perception, undermines trust in democracy, and polarizes communities [83] | Stringent ad-review policies, independent fact-checkers, ad libraries, and public awareness campaigns [84] |
| Fake News [85] | Erodes trust in information sources, threatens democracy, public safety, and societal harmony [86] | AI-driven fact-checking tools, partnerships with independent fact-checkers, digital literacy campaigns, and transparency features [87] |
| Deepfake Manipulation [88] | Creates highly realistic fabricated media used in misinformation, defamation, and political manipulation, eroding trust and distorting public perception [89] | Advanced detection algorithms (facial landmark detection, frequency analysis, deep learning), media literacy, and collaborative efforts among platforms, researchers, and policymakers [90] |
| AI-flenerated Misinformation [91] | Automates the creation of convincing fake news and misleading content, amplifying information disorder and overwhelming users with deceptive narratives [92] | AI-based detection models, fact-checking systems, platform-level interventions, and public awareness campaigns [87] |

Our categorization process, pivotal for social cybersecurity detection, begins with ML methods that learn from historical patterns to make future projections [93]. The second category, discrete models [94; 95], provides detailed frameworks for representing complex systems and interactions, crucial for understanding and detecting attack dynamics. Metaheuristic algorithms then address the complexities endemic to social cybersecurity, refining solutions to accurately counter threats. Lastly, Agent-based modeling (ABM) simulates multifaceted interactions on platforms like social media, making it invaluable for detecting emergent behaviors tied to cybersecurity threats.

The increasing use of AI in social cybersecurity detection necessitates advanced methodologies that account for adversarial tactics used to evade detection. One challenge is how attackers exploit AI-based defense mechanisms by leveraging deceptive strategies to manipulate detection models. For instance, research by Hong et al. (2024) highlights how expectancy violations in AI interactions can influence user trust, which has implications for adversarial deception tactics in cybersecurity [96]. To enhance detection accuracy, integrating adversarial machine learning techniques and incorporating insights from human-computer interaction research can improve model robustness against manipulation and evasion strategies.

Throughout this discussion, we reference literature to offer a holistic understanding of these methods' roles in forecasting and countering social cybersecurity threats.

## 4.1. Machine Learning in Social Cybersecurity

In this section, we focus on ML, the foremost category of methods employed in detecting social cybersecurity attacks. We explore a spectrum of approaches, from classifier and clustering models to other related technologies. An overview of these methods, including relevant research papers, is provided in Table 4 and the summary section.

ML techniques [93; 32; 97] are instrumental in detecting imminent social cybersecurity threats and understanding their trends. Classification models in ML categorize data, facilitating the early detection of threats, while regression and time series analysis utilize historical data to identify and anticipate attack patterns. The ML process encompasses a training stage, where models learn from past data, followed by a testing phase that focuses on detecting and understanding the nature of threats. The choice between supervised, unsupervised, or semi-supervised learning is dictated by the specific requirements in social cybersecurity, balancing the detection of current threats with insights into potential future behaviors.

Below, we'll examine the literature on detecting social cybersecurity attacks with ML models. We'll focus on three key categories: classifier models, clustering models, and time series, along with an exploration of text analysis techniques in this domain.

### Classifier Models

In the arena of Support Vector Machines (SVMs) [98], several standout studies have been conducted. Sabottke et al. [99] applied an SVM classifier to Twitter data, offering a novel method to predict real-world exploits, detecting them earlier than other techniques. Sharif et al. [100] leveraged

SVM to pinpoint extreme behaviors on Twitter, particularly in political discourse, and found it especially potent when paired with the bigram feature set. Benevenuto et al. [101] employed SVM classifiers on an expansive dataset, underscoring its efficacy in discerning Twitter spammers. El-Mawass et al. [102] blended SVM with a Markov Random Field (MRF) to detect spammers on Twitter, with a special emphasis on the utility of the SVM classifier that utilized the RBF kernel. Moreover, Mostafa et al. [103] showcased a technique using SVM to recognize spam campaigns on the platform.

Transitioning to the domain of Convolutional Neural Networks (CNNs) [104], various innovative approaches have been presented. Dionísio et al. [105] employed deep learning to identify cyber threats on Twitter, particularly security-related tweets about IT infrastructures. Shin et al. [106] harnessed a CNN classifier to gather cybersecurity intelligence from multiple sources on Twitter. Furthering the use of CNN, Alves et al. [107] highlighted the potential of Twitter as a hub for security alerts, while Simran et al. [108] underscored a CNN-GRU-Keras model's proficiency in detecting threats. Alguliyev et al. [109] combined CNN with an LSTM model to pinpoint DDoS attacks, and Huang et al. [110] illustrated CNN's potency in classifying emergency-related posts on Sina Weibo. Adding to the ensemble techniques, an ensemble CNN+Bi-GRU model was employed to tackle the challenge of COVID-19 misinformation on Twitter [111].

Focusing on Long Short-Term Memory (LSTM) networks [112], Wang and Zhang [113] zeroed in on DDoS attack detection on social media, praising the hierarchical LSTM model's commendable performance. In the Random Forest (RF) algorithm landscape [114], Maziku et al. [115] used RF for Twitter spam detection, while Fazil and Abulaish [116] emphasized the RF classifier's superiority in detecting Twitter spammers amidst class imbalances.

In exploring other classifier models within social cybersecurity, several research endeavors are noteworthy. Gupta et al. [117] utilized the Naïve Bayes (NB) approach [118] for Twitter spam detection and observed enhanced accuracy with a unified strategy. Sharif et al. [119], by leveraging the Stochastic Gradient Descent (SGD) method [120], successfully pinpointed suspicious Bengali tweets, emphasizing the compatibility of the SGD classifier with 'tf-idf'. Meanwhile, Le et al. [121] turned to the Nearest Centroid strategy [122] to extract cyber threat intelligence from Twitter, highlighting their classifier's robust performance. Finally, Zong et al. [123] integrated the Logistic Regression (LR) classifier [124] with a one-dimensional Convolutional Neural Network (CNN) for a nuanced Twitter threat assessment, innovating with a severity scoring model and spotlighting trustworthy warning accounts.

### Clustering Models

Shao et al. [125] embarked on a journey to unearth malicious cybercriminal activities on Twitter using the Gaussian Mixture Model (GMM) [126]. In their quest, they put several clustering algorithms to the test, ultimately recognizing the Expectation Maximization (EM)-based GMM—especially when synergized with a kernel filter—as the most adept at singling out malicious users. Taking the baton, Ritter et al. [127] turned their focus towards the Expectation Maximization (EM) method [128] to sift through Twitter for cyber attack-related events. By harnessing seed queries and weaving in NLP techniques, their approach unfurled impressive AUC values across various cyber attack classifications. Diving into a mammoth dataset of 14 million entries, Eshraqi et al. [129] employed the DenStream clustering algorithm, drawing a clear demarcation between spam and non-spam tweets. Their method's mettle was significantly reflected in the nuanced adjustments of the epsilon parameter. Pivoting to broader horizons, the SONAR framework [130] champions real-time detection, geolocation, and categorization of Twitter-centric cybersecurity incidents. This robust framework takes advantage of the nuances of Locality-Sensitive Hashing (LSH) [131] to make its mark in the field.

**Time Series**

Mahmood et al. [132] recognized the power of time series analysis [133; 134; 135] and harnessed it to anticipate phishing attacks on big-league brands, such as AOL and Facebook, delving deep into the PhishMonger dataset. Meanwhile, Gerber [136] elegantly amalgamated Twitter data with Chicago crime records, setting the stage for enhanced crime prediction—a move that underscored the prospect of optimized resource deployment. Goyal et al. [137] conducted a comprehensive analysis of various web signals, ranging from Twitter to the hidden areas of the dark web. Their goal was to accurately predict cyberattacks, emphasizing the importance of tailored monitoring strategies. This was especially evident during significant events such as Euro 2016, Javed et al. [138] channeled Twitter data, architecting a forward-looking system poised against drive-by download assaults. Drawing from the Hackmageddon Master List dataset, the insights in [139] brought to the fore the prowess of time series prediction techniques in the evaluation of cyber onslaughts. Not to be left behind, Potha et al. [140] ventured into the realm of cyberbullying detection, rigorously sifting through diverse feature extraction approaches and classifiers, all backed by data from Perverted-Justice.

**Text Analysis Techniques in Social Cybersecurity**

In the realm of social cybersecurity, researchers have employed various text analysis techniques. Mittal et al. [141] harnessed Named Entity Recognition (NER) to generate cybersecurity threat alerts from a dataset of English tweets. Vadapalli et al. [142] developed an Open Source Intelligence (OSINT) system that automatically detects and analyzes cybersecurity intelligence from Twitter. Chambers et al. [143] employed Latent Dirichlet Allocation (LDA) [144] and other NLP techniques to detect Distributed Denial of Service (DDoS) attacks from Twitter data, with a notable emphasis on their proposed PLDAttack model. Seed queries have been instrumental for researchers like Ritter et al. [127], Khandpur et al. [145], and Sceller et al. [130] in gathering event-specific information, with examples such as the Sarah Palin hacking incident. Khandpur et al. [145] further explored the potential of Query Expansion (QE) [146] in detecting cyber attacks using social media, introducing a Dynamic Query Expansion (DQE) algorithm for refined query selection. Lastly, Mukunthan et al. [147] utilized URL Blacklists to identify Twitter spam and delved into the analysis of user behavior patterns linked to spamming.

**Summary**

This section covered research already conducted in social cybersecurity using various ML approaches. Diverse techniques such as classifiers, clustering models, and other technologies have been employed for tasks including threat detection, malicious activity identification, spam classification, cybersecurity event monitoring, and information extraction from social media data. Within the field of social cybersecurity, particularly in the subdomain of detecting attacks, a conventional methodology mirroring standard ML practices are widely adopted. This methodology, depicted in Figure 4, encompasses a pipeline consisting of the following steps:

1. *Data Collection:* Gather relevant data specifically focused on historical, social cybersecurity attacks, including attack details, attack vectors, social media platforms targeted, and affected users. This data will serve as the foundation for training the detection model.

2. *Data Preprocessing:* Clean and preprocess the collected data, giving special attention to features that are indicative of social cybersecurity attacks. Handle missing values, outliers, and inconsistencies while preserving the integrity of attack patterns for accurate detection.

3. *Feature Extraction:* Extract features that have strong predictive power for social cybersecurity attacks. Consider incorporating domain-specific knowledge and expertise to identify relevant features, such as specific linguistic patterns used in social engineering attacks or behavioral indicators of malicious accounts.

4. *Dataset Splitting:* Split the preprocessed data into training and testing datasets, ensuring that the distribution of social cybersecurity attacks is properly represented in both sets. This enables reliable evaluation of the model's detection performance.

5. *Model Selection:* Choose ML algorithms that are effective in detection tasks and suitable for capturing temporal patterns in social cybersecurity attacks. Consider models like recurrent neural networks (RNNs), long short-term memory networks (LSTMs), or time-series forecasting algorithms.

6. *Model Training:* Train the selected model using the training dataset, emphasizing its ability to detect future social cybersecurity attacks. Enable the model to learn the underlying patterns and dynamics that contribute to accurate detections.

7. *Model Evaluation:* Evaluate the model's performance using evaluation metrics specifically designed for detection tasks in social cybersecurity. Assess its ability to identify new attack patterns, detect attacks within a given timeframe, and accurately identify emerging threats.

8. *Hyperparameter Tuning:* Fine-tune the model's hyperparameters with a focus on optimizing its detection capabilities for social cybersecurity attacks. Adjust parameters related to temporal modeling, regularization, and learning rate to enhance the model's performance.

9. *Detection:* Utilize the trained model to make detection on new, unseen data, specifically targeting the likelihood and occurrence of social cybersecurity attacks. Leverage the model's ability to capture temporal dynamics and emerging patterns for more accurate detections.

10. *Model Deployment:* Deploy the model in a real-world environment, integrating it into systems that monitor social media platforms and communication channels for proactive defense and early threat detection. Continuously analyze incoming data to provide timely detections of social cybersecurity attacks.

11. *Model Updating:* Regularly update and retrain the model with new data to adapt to evolving attack techniques and changing threat landscapes. Incorporate feedback mechanisms and continuous learning to improve the model's detection capabilities over time.

By refining the steps to focus on the detecting aspects, you can better align the process with the specific goal of anticipating and mitigating social cybersecurity attacks.

## 4.2. Discrete Models in Social Cybersecurity

This section focuses on discrete models used in social cybersecurity attack detection. We explore graph models such as attack graphs, Markov models, and Bayesian networks, as well as the application of game theory principles. Table 5 and the summary section offer an overview of the methods discussed and relevant research papers in this category.

**Graph Models**

In this section, we will explore various graph models, including attack graphs, Markov models, and Bayesian Networks, and their application in the context of detecting social cybersecurity attacks.

Attack graphs [148; 149; 150] offer visual insights into potential attack paths, especially within social platforms. Markov models [151; 152; 153; 154; 155] capture the sequential dynamics of attacks, providing insights based on historical data and aiding in anticipatory measures. Bayesian Networks (BNs) [156] model the interdependencies among critical factors, using probabilistic reasoning to handle uncertainties. Together, these methods form a comprehensive approach to understanding, predicting, and strategizing against threats in social cybersecurity.

Below, we will review the literature on detecting social cybersecurity attacks using attack graphs, Markov Models, and Bayesian Networks.

In the realm of attack graphs, several noteworthy studies stand out. Chen et al. [157] utilized a CVE-Author-Tweet (CAT) graph derived from Twitter data to predict the exploitation of Common Vulnerabilities and Exposures (CVEs). Feng et al. [158] introduced the BotRGCN model for Twitter, leveraging a Relational Graph Convolutional Network (RGCN). Gao et al. [159] employed clustering and graph theory techniques to detect social spam campaigns on Facebook, emphasizing the identification of potential spam based on user posts. Lastly, Lippmann et al. [160] endeavored to pinpoint malicious cyber discussions across platforms such as Twitter, Stack Exchange, and Reddit by harnessing Human Language Technologies (HLT) and constructing various graphs to depict the communication dynamics.

In the domain of Markov models, several studies have made significant contributions. El-Mawass et al. [161] concentrated on detecting spammers on Twitter, examining a dataset that covered 767 users across four unique categories. They relied on symmetric and asymmetric Markov Random Fields (MRFs) for their classification. Aleroud et al. [162], aiming to identify pro-ISIS accounts on Twitter, analyzed datasets linked to the 2015 Paris terrorist incidents and tweets with ISIS-associated keywords. Their methodology combined a Markov Chain (MC) with a Topic Model (TM) and an SVM classifier. Further, Li et al. [163] worked on recognizing health campaign promoters on Twitter, particularly those championing anti-smoking campaigns. They adopted the Typed Markov Random Field (T-MRF) algorithm for their classification process. Rounding out the list, Qiao et al. [164] applied a Hidden Markov Model (HMM) with the goal of predicting social unrest events, sourcing their data from a segment of the Global Database of Events, Language, and Tone (GDELT) project.

In the realm of Bayesian Networks, significant contributions can be found in the literature. Okutan et al. [165] embarked on the task of developing a predictive model for a spectrum of cyber attack types—including malware, scan, defacement, malicious email, malicious URL, and Denial of Service (DoS). Their approach hinged on datasets sourced from GDELT, Twitter, and documented cyber incidents, all processed through Bayesian Networks (BNs). Another research [166] tapped into data from Twitter, Hackmageddon, and GDELT.

Their goal was to engineer a Bayesian Network (BN) tailored for the detection of cyber attacks, with a special emphasis on types like defacement, Denial of Service (DoS), and malicious emails or URLs.

**Game Theory**

In this section, we examine the use of the Game Theory approach in the detection of social cybersecurity attacks.

Game theory [167; 168; 169; 170; 171] offers a mathematical approach to analyze strategic interactions between attackers and defenders in the realm of social cybersecurity. It captures the attacker-defender dynamics, considering the strategic decisions, trade-offs, risks, and rewards inherent in these engagements. Through game theory, researchers can predict likely attack vectors, ascertain optimal defense strategies, and estimate the outcomes of cybersecurity confrontations. This enables effective resource allocation, prioritized defense measures, and the crafting of proactive security tactics.

Below, we'll review the literature on detecting social cybersecurity attacks using game theory.

Griffin and Squicciarini [172] explored user behavior related to deception in social media by proposing a game theoretical model to analyze user tendencies in identity disclosure influenced by peer behavior. Kamhoua et al. [173] introduced a game theoretical approach to guide users in social networks towards an optimal data-sharing policy, weighing the conflicting interests between users and attackers. Liang et al. [174] used game theory to devise an optimal data-forwarding strategy aimed at preserving privacy in mobile social networks. Mohammadi et al. [175] utilized a signaling game in social networks to differentiate between regular users and attackers, strategically deploying fake avatars to bait and identify malicious intents, drawing inspiration from concepts in [176; 177]. In [178], community detection in online social networks was approached using an algorithm rooted in game theory. The study in [179] presented a game theoretical approach to analyzing attacks on social network services, emphasizing the dynamics of user-controlled data sharing. [180] developed a game theoretic model for online social networks, focusing on the balance between information sharing and security. Finally, Zhao et al. [181] introduced a game theoretic framework to study the dynamics of colluders and the cooperation amongst attackers in multimedia social networks, especially concerning multimedia fingerprinting and unauthorized content usage during collisions.

**Summary**

This section presents an overview of the research conducted in social cybersecurity utilizing discrete models, including graph attacks, Markov models, Bayesian networks, and game theory. These models play a crucial role in detecting social cybersecurity attacks. By capturing state transitions and behavioral patterns, they facilitate the analysis of social behavior within the security context. The pipeline for leveraging these models involves several key steps, as illustrated in Figure 5. These steps include:

1. *Data Collection:* Gather relevant data from communication channels targeted in social cybersecurity attacks. This includes email, social media, and instant messaging platforms. The collected data should encompass both benign and malicious activities to train the discrete models effectively.

2. *Data Preprocessing:* Clean and transform the collected data to ensure its quality and suitability for analysis with discrete models. Handle missing values, correct inconsistencies, and standardize the data to maintain its integrity during preprocessing.

3. *Feature Extraction:* Extract features from the preprocessed data that have strong detective power for social cybersecurity attacks. Consider features such as specific phrases, sentiment analysis scores, communication frequency, network-based features, or other elements indicative of malicious activity. These features serve as inputs to the discrete models.

4. *Modeling:* Develop and train discrete models such as graph attacks, Markov models, or Bayesian networks using the preprocessed data and extracted features. Graph attacks can model social networks and identify suspicious behaviors within them. Markov models can capture transitions between different communication patterns, while Bayesian networks can model the probabilistic relationships between different features.

5. *Detection:* Utilize the trained discrete models to detect potential social cybersecurity attacks. Analyze current behavior patterns and compare them to the models to identify possible threats and make informed detections about the type of attack, potential targets, and likely timing. Each discrete model can offer unique insights into the detection process.

6. *Evaluation:* Assess the accuracy and effectiveness of the discrete models in detecting social cybersecurity attacks. Compare the models' predictions with real-world outcomes and utilize appropriate evaluation metrics to measure their performance. This evaluation helps identify areas for improvement and fine-tuning of the models.

7. *Model Deployment:* Deploy the trained discrete models in a detection environment that can handle real-time data processing and classification. This could involve deploying the models on cloud servers with appropriate resources or on edge devices for real-time analysis and response to social cybersecurity threats.

8. *Updating Model:* Regularly update the discrete models with new data as social engineering tactics evolve. Incorporate mechanisms to continuously retrain and update the models, ensuring their relevance and effectiveness in detecting social cybersecurity attacks over time.

By incorporating discrete models into the steps and considering their specific capabilities, the process becomes more focused on the detection task for social cybersecurity attacks. This approach allows for a nuanced analysis of social behavior within the security context and facilitates improved prediction accuracy and proactive threat detection.

Table 4: A summary of detection methods. Approaches based on machine learning.

| Ref. | Objective | Approach/Model | Dataset | Limitation | Year |
|------|-----------|----------------|---------|------------|------|
| [105] | Detect cyber threats | NER, SVM, MLP, CNN+random, CNN+GloVE, CNN+Word2Vec | Twitter | Trained on a limited dataset | 2019 |
| [127] | Extract computer security events | Seed queries, SVM, EM, NER, LR | Twitter | Inadequate performance due to non-random sampling of seeds | 2015 |
| [145] | Detect cyber attacks | Seed queries, AP, NER, DQE | Twitter | Low precision and recall for DoS attacks and account hijacking | 2017 |
| [141] | Generate alerts for cybersecurity threats | NER | Twitter | Incorrect discarding of tweets due to various factors | 2016 |
| [99] | Predict real-world exploits | SVM | Twitter | Security systems without the need for secrets or confidential information | 2015 |
| [130] | Automatically detect cybersecurity events | LSH | Twitter | Inability to identify the target and source of an attack | 2017 |
| [106] | Detect cybersecurity intelligence on Twitter | CNN, RNN, LSTM | Twitter | No significant weaknesses | 2020 |
| [142] | Automatically detect and analyze cyberse- curity intelligence | NER | Twitter | Limited to a single OSINT source (Twitter) | 2018 |
| [182] | Cyberthreat detection | CNN, RNN, LSTM, NER | Twitter | Trained on a limited or small-scale dataset | 2020 |
| [107] | Provide evidence of timely and impactful security alerts on Twitter | CNN | Twitter | Possible failure to capture additional cases, potential human error in manual data processing | 2020 |
| [108] | Detect cybersecurity threats | SVM, CNN, DNN, RNN, GRU, fastText | Twitter | No significant weaknesses | 2019 |
| [121] | Gather cyber threat intelligence using novelty classifiers | SVM, NC | Twitter | Lack of specific Named-Entity Recognition (NER) phase for identifying entities related to vulnerabilities | 2019 |
| [123] | Analyze severity of cybersecurity attacks | LR, CNN | Twitter | Low severity scores for high severity threats lacking detailed tweet contents | 2019 |
| [143] | Detect DDoS attacks | LDA, LR | Twitter | Low recall, manual monitoring required for identifying false positives | 2018 |
| [115] | Spam detection | RF | Twitter | Longer classification process compared to existing solutions | 2020 |
| [117] | Spam detection | NB, DT, unsupervised clustering | [101] | Poor performance of DT for non-spam accounts | 2015 |
| [147] | Analysis of spam and user behavior | URL blacklist | Twitter | Difficulty detecting spam using URL shorteners like Bitly | 2021 |
| [125] | Monitor and detect malicious activity of cybercriminals | Kernel Filter, EM-based GMM, K-Means, hierarchical agglomerative clustering | Twitter | No significant weaknesses | 2019 |
| [101] | Spam detection | SVM | Twitter | No significant weaknesses | 2010 |
| [129] | Spam detection | DenStream Clustering | Twitter | No significant weaknesses | 2015 |
| [113] | Detect DDoS Attacks | SVM, LSTM, NER | Twitter | No significant weaknesses | 2017 |
| [109] | Detect DDoS Attacks | CNN, LSTM | Twitter | No significant weaknesses | 2019 |
| [102] | Spam detection | Similarity approach with SVM and MRF | Twitter | No significant weaknesses | 2018 |
| [110] | Classify emergencies | CNN, kNN, DT, NB, SVC | Sina Weibo | Use of seed words may result in missing relevant tweets, highly imbalanced dataset, reliance on a single data source | 2010 |
| [103] | Spam campaign detection | SVM | Twitter | System does not consider temporal similarity of post timings | 2020 |
| [119] | Detect suspicious tweets | RF, NB, DT, SGD, LR | Twitter | Low accuracy | 2020 |
| [100] | Identify extreme behavior | RF, NB, DT, SVM, kNN, bagging, boosting | Twitter | System does not incorporate semantic analysis in processing | 2019 |
| [183] | Detect crime | RF | AboutIsis | System does not utilize associated images and videos in tweets | 2019 |
| [116] | Spam detection | BN, DT, RF | Twitter | No significant weaknesses | 2018 |
| [132] | Phishing attack forecasting | MLP, Prophet, Linear LSTM, ARIMA, linear regression, RF | PhishMonger project | No obvious weaknesses | 2020 |
| [136] | Crime detection | LDA, LR | Chicago crime data | Limited analysis of tweet structure and temporal effects | 2014 |

| [137] | Cyber attack signal discovery | ARIMA, ARIMAX, LSTM, GRU | Dark web, Twitter, blogs, vulnerability DB, honeypots | No obvious weaknesses | 2018 |
|---|---|---|---|---|---|
| [138] | Drive-by download attack prediction | NB, Bayes Net, J48, MLP | Twitter | Potential failure in detecting short-lived cyber-criminals | 2019 |
| [139] | Cyber attack prediction through text analysis | NER, Classifier | Hackmageddon Master List | Lack of multiple reports for the same attack in the dataset | 2018 |
| [140] | Cyberbullying detection | MLP, SVM | Perverted-Justice dataset | Absence of victim's response to cyberbullying messages | 2014 |
| [111] | Fake news detection | CNN+Bi-GRU | Twitter | The research's focus on COVID-19 tweets may limit its applicability to broader misinformation contexts | 2022 |

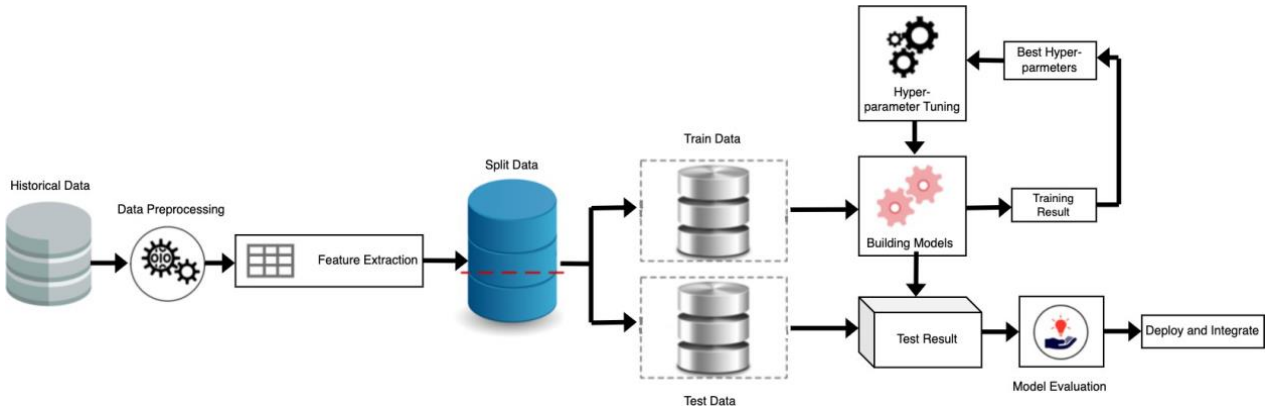

**Figure 4:** The ML system design starts with raw data collection and ends with selecting trustworthy ML models for detecting social cybersecurity attacks.

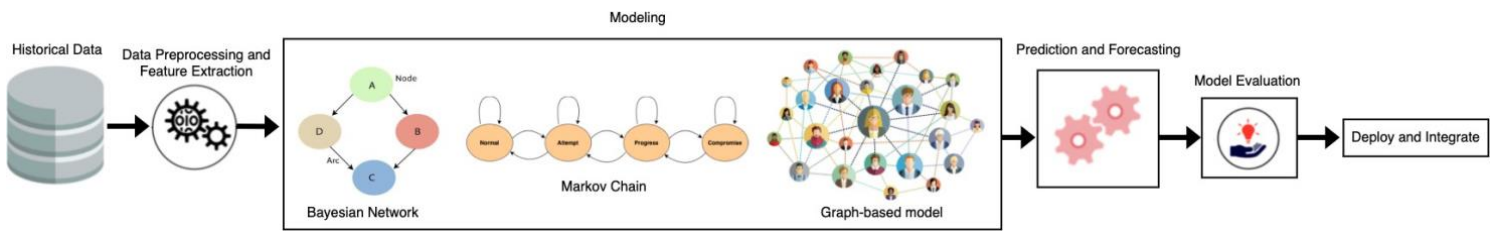

**Figure 5:** The discrete model design starts with raw data collection and ends with model evaluation for detecting social cybersecurity attacks.

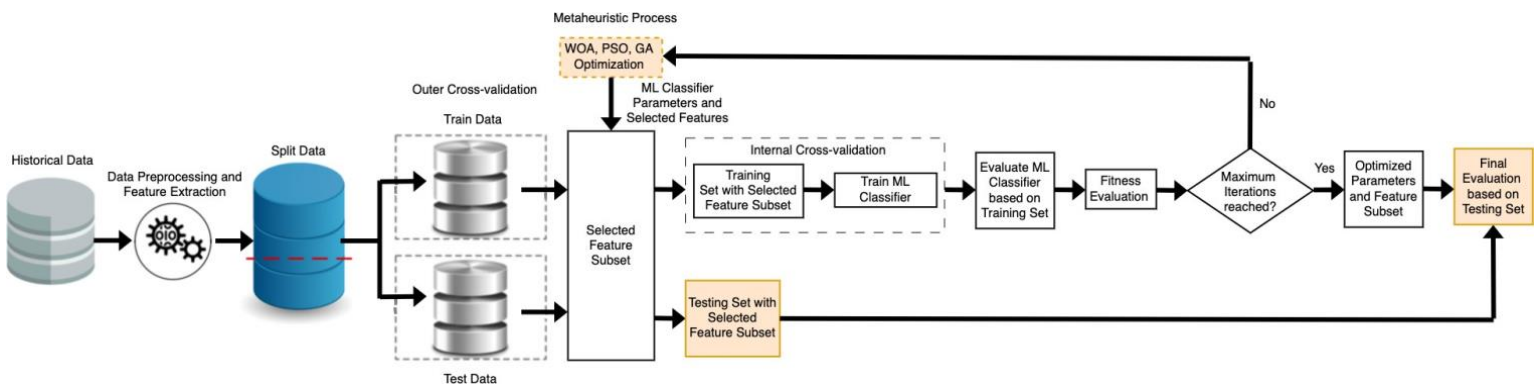

**Figure 6:** The ML system with a metaheuristic algorithm starts with raw data collection and ends with model evaluation for detecting social cybersecurity attacks.

**Table 5**
A summary of detection methods. Approaches based on discrete models.

| Ref. | Objective | Approach/Model | Dataset | Limitation | Year |
|------|-----------|----------------|---------|------------|------|
| [161] | Spam detection | MRF | Twitter | Additional models evaluated for improved accuracy in real-world applications | 2020 |
| [157] | Detecting vulnerability exploitation via Twitter | Multi-layered graph | Twitter (Limited) | Incorporation of cross-platform dataset to predict vulnerabilities | 2019 |
| [160] | Detecting malicious cyber discussions | flraph database | Twitter, Stack Exchange, Reddit | Limited cross-domain performance | 2015 |
| [158] | Bot account detection | Relational graph convolutional network | Twitter | No obvious weaknesses | 2021 |
| [159] | Detecting social spam campaigns | Clustering, graph theory | Facebook | Uncertain effectiveness in enticing users to click malicious URLs | 2010 |
| [162] | Identifying terrorist accounts | MC, RAF, kNN, and SVM | Twitter | Focus on English tweets and periodic retraining | 2020 |
| [163] | Detecting promoters of campaigns | T-MRF | Twitter | No obvious weaknesses | 2014 |
| [164] | Detecting events of social unrest | HMM | flDELT | Inability to differentiate between widespread and localized news coverage | 2017 |
| [165] | Detecting cyber-attacks | BN | flDELT, reported cyber incidents, Twitter | No obvious weaknesses | 2017 |
| [166] | Detecting cyber-attacks | BN | flDELT, Twitter, Hackmageddon | Inaccurate long-range predictions | 2017 |
| [172] | Modeling deception in social media | flame theory | Survey data | Need for further research on user actions and impact of outcomes on social image | 2012 |
| [173] | Optimal data sharing on social networks | flame theory, two-player zero-sum Markov game | None | No obvious weaknesses | 2012 |
| [174] | Privacy-preserving data forwarding on mobile social networks | flame theory | None | No obvious weaknesses | 2012 |
| [175] | Analysis of deception in social networks | flame theory, signaling game | None | No obvious weaknesses | 2016 |
| [178] | Community detection | flame theory | Karate, Dolphins, Football, Polbooks | Limited results and slow run-time without parallelism | 2019 |
| [179],[180] | Analysis of social network services | flame theory | None | Need for further research on honeytokens, deployment strategies, and distribution monitoring | 2014, 2013 |
| [181] | Analyzing user dynamics in social networks | flame theory | None | No obvious weaknesses | 2012 |

## 4.3. Metaheuristic Algorithms in Social Cybersecurity

This section delves into the application of metaheuristic algorithms in detecting social cybersecurity attacks. We examine methodologies that belong to two categories: trajectory-based and population-based metaheuristics. The latter encompasses concepts derived from nature, such as biological and swarm-based approaches, as well as those grounded in evolutionary theory. To gain a comprehensive understanding, Table 6 presents a comprehensive compilation of the methods discussed in this segment, along with the corresponding scholarly articles. Further insights can be found in the subsequent summary section.

Metaheuristic algorithms [196; 197; 198], including genetic algorithms, particle swarm optimization, and others, offer robust methods for optimizing prediction models in social cybersecurity. By efficiently exploring the solution space, these algorithms refine prediction models, enhancing their accuracy in detecting attacks [199; 200; 201; 202]. They excel in processing vast datasets, unveiling hidden patterns crucial for threat identification [203; 204; 205]. Consequently, metaheuristics assist researchers in decision-making and proactive defense formulation.

Among these, the Whale Optimization Algorithm (WOA) [206], inspired by humpback whales' hunting behavior, stands out. Particularly adept at feature selection [207; 208], WOA optimizes cybersecurity prediction models, bolstering their efficacy. Its unique attributes and minimal parameter adjustments make it an indispensable tool in refining social cybersecurity strategies.

Below, we'll review the literature on detecting social cybersecurity attacks using metaheuristic algorithms.

In the burgeoning field of social media analysis integrated with metaheuristic algorithms, myriad noteworthy studies have emerged. For instance, Sánchez-Oro and Duarte [184] drew a comparison between the Multi-Start Iterated Greedy (MSIG) method and the Ant Colony Optimization (ACO) algorithm, focusing on community detection. On the other hand, Sangwan and Bhatia [185] delved into rumor detection, utilizing the Wolf Search Algorithm (WSA) and Decision Trees (DT) to analyze comments from renowned global personalities.

Cyberbullying, a pressing concern in today's digital age, saw innovative approaches from Singh et al. [186] and Al-Ajlan et al. [187]. While Singh and his team combined the Cuckoo Search Algorithm (CSA) with a Support Vector Machine (SVM) for detection, Al-Ajlan and collaborators harnessed an insect behavior-inspired algorithm and paired it with a Convolutional Neural Network (CNN).

Shifting the lens to spam detection, Al-Zoubi et al. [188] differentiated between spam and non-spam tweets using a blend of algorithms, and Aswani et al. [189] employed the Levy flight Firefly Algorithm (LFA) alongside K-means clustering for Twitter spam detection. Tackling other security threats, Baydogan and Alatas [190] focused on hate speech classification, and Villar-Rodriguez et al. [191]

**Table 6**
A summary of detection methods. Approaches based on metaheuristic algorithms.

| Ref. | Objective | Approach/Model | Dataset | Limitation | Year |
|------|-----------|----------------|---------|------------|------|
| [184] | Community detection | Integrated flreedy Algorithm | Facebook, Twitter | No obvious weaknesses | 2018 |
| [185] | Rumor detection | WSA, DT | PHEME dataset, Twitter/Instagram | Low classification accuracy | 2020 |
| [186] | Cyberbully detection | CSA, SVM | Twitter, ASKfm, FormSpring | No obvious weaknesses | 2020 |
| [187] | Cyberbully detection | Insect-inspired metaheuristic algorithm, CNN | Twitter | Does not support Arabic text | 2018 |
| [188] | Spam profile detection | WOA, PSA, flA, SVM | Twitter | No obvious weaknesses | 2018 |
| [189] | Spammer detection | LFA, k-means clustering | Twitter | Content and semantic analysis impacted by satire and slang. Ignores useful features from user profiles. Does not examine shared links in-depth | 2018 |
| [190] | Hate speech detection | ALO, MFO, SSO, TSA, kNN, DT, SMO, MCC, J48, NB, RF, Ridor | Twitter | Metaheuristic runtime increases with dataset size and class imbalance. Multiple runs required to determine optimal parameters | 2021 |
| [191] | Impersonation attack detection | Bio-inspired metaheuristic algorithm, k-means clustering | WOSN 2009 conference Facebook user repository | No obvious weaknesses | 2017 |
| [192] | Cybercrime detection | CSA, SVM, NB | Twitter, ASKfm, FormSpring | Framework implemented for specific classifiers | 2019 |
| [193] | Link prediction | PSO, SVM | Twitter | No obvious weaknesses | 2020 |
| [194] | Face spoofing detection | PSA, ACO, SA, SVM | NUAA PID dataset | No obvious weaknesses | 2020 |
| [195] | Spam detection | PSO, SA, ACO, DE, SVM, DT | Facebook | No obvious weaknesses | 2017 |

targeted impersonation attacks on social platforms using a bio-inspired approach.

Other innovative applications include Singh and Kaur's [192] use of the Cuckoo Search Algorithm (CSA) for broader social media cybercrime, Rezaeipanah et al.'s [193] employment of Particle Swarm Optimization (PSO) for predicting social network connections, and Yılmaz et al.'s [194] incorporation of PSO for feature selection in face spoofing detection.

In the context of Facebook spam detection, Sohrabi and Karimi [195] tapped into a suite of metaheuristic algorithms, prominently PSO. However, a notable void exists in harnessing evolutionary-based algorithms for predicting social cybersecurity threats, and signposting promising directions for future inquiries.

**Summary**

This section of the research paper focused on existing studies conducted in social cybersecurity using metaheuristic algorithms for detecting social cybersecurity attacks. Both trajectory-based and population-based metaheuristics were examined, with the population-based category encompassing nature-inspired techniques such as bio and swarm algorithms and evolutionary-based approaches. It is important to note the identified research gap regarding the lack of studies on evolutionary-based methods. In the field of social cybersecurity research, a common approach was observed, which aligns with the typical utilization of metaheuristic algorithms. This approach is illustrated in Figure 6. The associated pipeline involves the following steps:

1. *Data Collection:* Gather data specifically related to social cybersecurity attacks, focusing on areas commonly targeted, such as emails, social media platforms, and instant messaging services. This ensures that the collected data is relevant to the detection task.

2. *Data Preprocessing:* Emphasize the cleaning and transformation of the collected data to maintain its integrity and suitability for analysis with metaheuristic algorithms. Address missing data and outliers and normalize the data to ensure consistency and improve the algorithms' performance.

3. *Feature Extraction:* Identify and extract features that are highly indicative of social cybersecurity attacks. These features could include specific phrases, sentiment analysis scores, communication frequency, network-related elements, or other relevant attributes that capture the essence of malicious activity.

4. *Problem Formulation:* Formulate the detection problem with a clear focus on social cybersecurity attacks. Define the objective function, constraints, and decision variables specific to detect these attacks. This ensures the metaheuristic algorithm is aligned with the desired outcomes.

5. *Algorithm Selection and Parameter Tuning:* Choose a suitable metaheuristic algorithm, such as Genetic Algorithms, Particle Swarm Optimization, or Simulated Annealing, based on the problem formulation and complexity. Optimize the algorithm's parameters to enhance its ability to forecast and predict social cybersecurity attacks effectively.

6. *Dataset Partitioning:* Split the dataset into training and testing sets, using techniques like cross-validation, to evaluate the model's performance robustly. This allows for accurate assessment of unseen data and helps avoid overfitting.

7. *Modeling and Fitness Evaluation:* Develop and train a model using the selected metaheuristic algorithm and the preprocessed data. Train an ML classifier or regression model using the training set and evaluate its fitness using the algorithm's objective function. This assesses the model's performance in detecting social cybersecurity attacks.

8. *Optimization:* Continuously optimize the feature selection process and tune the parameters of the ML model using the metaheuristic algorithm. This iterative optimization process aims to identify patterns, relationships, and optimal configurations that enhance the model's detection capabilities.

9. *Detection:* Utilize the optimized model to forecast and predict social cybersecurity attacks using the testing set. The trained model analyzes the current state, patterns, and features to provide insights into potential threats and their likelihood within the specified timeframe.

10. *Evaluation:* Compare the model's detections against actual outcomes to assess its performance in forecasting and predicting social cybersecurity attacks. Analyze the model's accuracy, precision, recall, F1 score, or other relevant metrics to measure its effectiveness and identify areas for improvement.

11. *Model Deployment:* Deploy the trained and evaluated model for real-world use, considering the specific requirements and constraints of the deployment environment. This may involve integrating the model into existing security systems or deploying it on cloud servers or edge devices for real-time predictions and proactive defense measures.

12. *Model Updating:* Regularly update the model with new data as social engineering tactics evolve and new attack patterns emerge. This ensures the model remains accurate, adaptable, and effective in detecting social cybersecurity attacks in dynamic threat landscapes.

By incorporating these considerations, the steps become more closely aligned with the detection task for social cybersecurity attacks using metaheuristic algorithms. This approach leverages optimization techniques and ML models to improve accuracy, enable proactive defense measures, and enhance overall cybersecurity posture.

## 4.4. Agent-based Modeling in Social Cybersecurity

In this section, we thoroughly investigate the use of Agent-based Modeling in detecting social cybersecurity attacks. Table 7, along with the summary section, provides a synopsis of the methods examined and the pertinent research papers in this particular field.

Agent-based modeling (ABM) [209; 210; 211] offers a computational simulation method to understand systems via individual agent interactions. In detecting social cybersecurity attacks, ABM captures the nuances of social behaviors, enabling simulations of users, attackers, and defenders in online platforms [212; 213]. Through this, it unveils emergent properties and patterns from these interactions, facilitating analysis of attack spread, defense effectiveness, and cybersecurity dynamics. This helps researchers detect attacks, assess varied scenarios, and devise proactive defense strategies.

Below, we'll review the literature on detecting social cybersecurity attacks using agent-based modeling.

Serrano et al. [214] propose an agent-based social simulation model to detect rumors on social media, contrasting it with a baseline model. Beskow et al. [215] introduces an agent-based model, Twitter sim, to investigate bot disinformation tactics on Twitter and delve into emergent behaviors like supporting key influencers. Onuchowska [216] uses agent-based modeling to analyze malicious behaviors on social media with the intent to mitigate the influence of malicious actors. Tseng and Nguyen [217] apply agent-based modeling to simulate rumor propagation on social media, highlighting its fidelity in portraying the dynamics of rumor spread. In [218], agent-based modeling is utilized to assess countermeasures against misinformation on social media, particularly those that curtail misinformation while promoting true information dissemination. Finally, Calay et al. [219] employ Agent-Based Modeling to scrutinize the impact of team formation strategies on cybersecurity team performance, culminating in the Collaborative Cyber Team Formation (CCTF) framework that offers a comprehensive understanding of cyber team dynamics.

### Summary

This section provides an overview of the existing research conducted in social cybersecurity using agent-based modeling for detecting social cybersecurity attacks. While spam detection has received considerable attention, other areas like crime and cyber-attack detection remain less explored. The full potential of agent-based modeling in social cybersecurity is yet to be realized, as comprehensive research in this area is still lacking. Figure 7 illustrates the typical approach followed in applying agent-based modeling in social cybersecurity research. The associated pipeline involves the following steps:

1. *Problem Definition:* Clearly define the problem you want to address using agent-based modeling in the context of detecting social cybersecurity attacks. This step focuses the modeling effort on the specific objective of detection.

2. *Identify Agents:* Identify the relevant agents involved in the social cybersecurity ecosystem, such as individuals, organizations, attackers, defenders, etc. This step ensures that the model includes all the necessary entities to accurately represent the social dynamics of cybersecurity.

**Table 7**
A summary of detection methods. Approaches based on agent-based modeling.

| Ref. | Objective | Approach/Model | Dataset | Limitation | Year |
|------|-----------|----------------|---------|------------|------|
| [214] | Detect spread of rumors | ABM | Twitter | No obvious weaknesses | 2015 |
| [215] | Determine bot maneuvers in spreading misinformation | ABM | Twitter | Cannot validate believably | 2019 |
| [216] | Detect malicious behavior | ABM | Twitter | Limited to Spain, Iran, Russia, and Venezuela. Incomplete communication modeling. Preferential attachment for relationship formation | 2020 |
| [217] | Rumor detection | ABM | N/A | Limited interaction modeling. Not a hybrid model. No specific simulation period | 2020 |
| [218] | Develop countermeasures against misinformation spread | ABM | Twitter | Non-realistic population size. Verification lacking. Limited datasets | 2021 |
| [219] | Examined team formation's effect on cybersecurity and introduced CCTF framework | ABM | N/A | The study uses Agent-Based Modeling without empirical validation, limiting real-world applicability | 2023 |

3. *Agent Behaviors and Interactions:* Define the behaviors and interaction rules of the agents based on real-world observations and expert knowledge of social cybersecurity attacks. Specify how agents communicate, exchange information, launch attacks, defend against attacks, and make decisions related to cybersecurity. These rules and behaviors shape the model's detection capabilities.

4. *Agent Attributes and Data:* Determine the attributes or variables associated with each agent that are relevant to detecting social cybersecurity attacks. Consider attributes such as susceptibility to attacks, awareness level, security measures in place, historical attack data, or any other factors that affect the likelihood and impact of attacks.

5. *Model Design and Implementation:* Design and implement the agent-based model using appropriate software or simulation platforms. Utilize tools like Mesa (the agent-based modeling in Python 3+), NetLogo [220], Repast [221], or MASON [222]. to create the environment for agent-based modeling. Implement the rules, mechanisms, and algorithms that govern agent behaviors, interactions, and decision-making processes specific to social cybersecurity attacks.

6. *Validation and Calibration:* Validate the agent-based model by comparing its outputs with real-world data or known scenarios. Calibrate the model parameters to ensure they accurately represent the observed behaviors and dynamics of the social cybersecurity system. This step ensures the model's accuracy and reliability in detection.

7. *Experimentation and Sensitivity Analysis:* Conduct experiments using the agent-based model to simulate different scenarios and observe the effects of various factors on the occurrence and spread of social cybersecurity attacks. Perform sensitivity analysis to understand the model's sensitivity to changes in input parameters and identify influential factors.

8. *Detection:* Utilize the agent-based model to detect social cybersecurity attacks based on different scenarios and inputs. Analyze the model's outputs to understand the potential impact and likelihood of future attacks, considering the dynamic interactions and behaviors of the agents.

9. *Evaluation and Interpretation:* Evaluate the performance of the agent-based model by comparing its detection with real-world events and data. Interpret the findings to gain insights into the dynamics of social cybersecurity attacks, identify potential preventive measures, and understand the effectiveness of different strategies. This evaluation ensures the model's effectiveness in detection.

10. *Model Refinement and Iteration:* Refine and improve the agent-based model based on feedback, additional data, and new insights. Iterate through the steps to enhance the accuracy and effectiveness of the model in detecting social cybersecurity attacks. Update the model to adapt to evolving social engineering tactics and incorporate new findings.

By incorporating these considerations, the steps become more closely aligned with the detection task for social cybersecurity attacks using agent-based modeling. This approach allows researchers to gain valuable insights into the dynamics of social cybersecurity threats, simulate different scenarios, and develop proactive strategies for prevention and mitigation.

## 5. Social Cybersecurity Tools and Public Datasets

In the field of social cybersecurity, the right tools and access to relevant datasets are pivotal for both research and practical applications. This section introduces both vital components. Initially, we examine prominent tools in social cybersecurity and social network analysis, helping readers make well-informed decisions based on literature reviews and expert feedback. Subsequently, we outline publicly available datasets, emphasizing their sources and utility in the realm of social cybersecurity.

### 5.1. Social Cybersecurity Tools

This section aims to provide a comprehensive comparison of the popular tools employed in the domains of social cybersecurity and social network analysis. Through a meticulous examination of the characteristics and functionalities of these tools, researchers and practitioners can make well-informed decisions when opting for appropriate tools for visualization and analysis. The evaluations are based on a comprehensive literature survey [223; 224; 225; 226] and expert opinions, ensuring the reliability of the insights presented here. For a concise overview, Table 8, accompanied by the summary section, offers a synopsis of the surveyed tools in this section, facilitating easy reference and understanding.

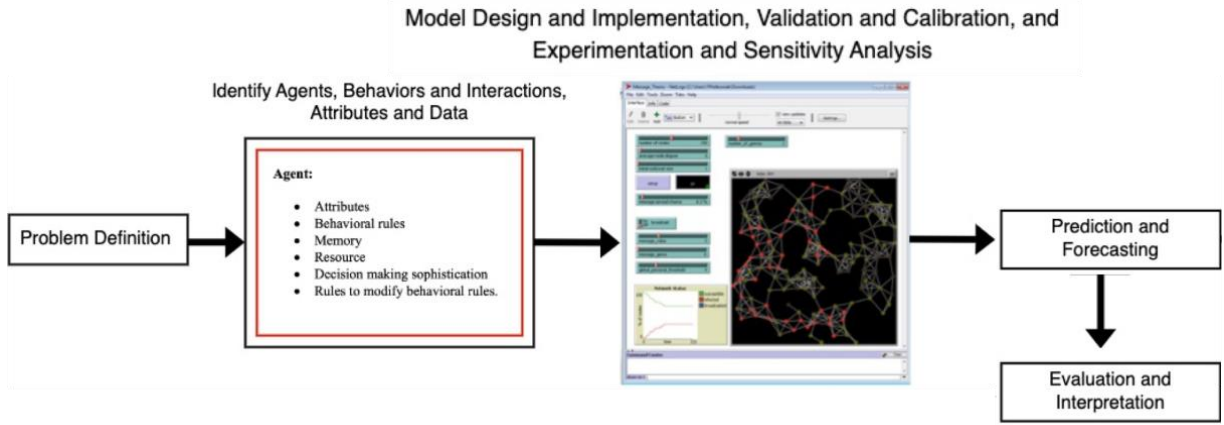

**Figure 7:** The agent-based modeling design starts with problem definition and ends with evaluation and interpretation for detecting social cybersecurity attacks.

**Desktop and Mobile Applications**

Pajek [227] is designed for social network analysis and visualization, aiding in identifying patterns and connections pertinent to cybersecurity threats within social networks. Gephi [228] provides capabilities for visualizing and manipulating dynamic graphs, facilitating the tracking of the evolution of cyber threats. AutoMap [229], specializing in text mining and analysis, is apt for extracting and interpreting cybersecurity themes from unstructured text on social platforms, and it integrates with ORA-LITE [230] for visualization and statistical analysis. UCINET [231] offers tools for data manipulation and analysis, crucial for understanding extensive datasets associated with social cybersecurity incidents. SocNetV [232] focuses on the visualization and analysis of social networks, allowing experts to delineate cyber threat networks. CFinder [233] is instrumental in discovering and visualizing communities within graphs, giving insights into community-centric cybersecurity issues. Concluding, Hootsuite [234] merges metrics from diverse social media platforms, providing a unified view of cybersecurity trends and anomalies across platforms.

**Libraries, APIs, and Plugins**

NetworkX [235] is tailored for creating and visualizing complex networks, which can be instrumental in visualizing cyber threat landscapes on digital platforms. The igraph library [236] provides extensive tools for graph creation and analysis, pivotal for understanding cyber interactions and potential vulnerabilities in the cyber realm. JUNG [237] serves as an API for modeling and analyzing networks, which can be beneficial in cybersecurity for pattern recognition and anomaly detection. NodeXL [238] focuses on social connections and interactions, making it pertinent for analyzing cybersecurity threats within social media ecosystems. JGraphT [239] offers functionalities crucial for visualizing cyber interactions and potential threat points in the digital realm. The Twitter API [240], by fetching real-time data from Twitter, can be employed to monitor cybersecurity threats and information dissemination in real-time on this platform. TensorFlow [241] and PyTorch [242] are both pivotal for developing machine learning classifiers, enabling threat intelligence and predictive cybersecurity through data analysis. Lastly, MuxViz [243] provides a framework to understand multilayer social networks, shedding light on complex cyber interactions across various platforms, thus assisting in multi-platform cybersecurity assessment and mitigation [244].

**Programming Languages**

R [245], with its comprehensive libraries like igraph and SNA, serves as a potent tool for data analysis and visualization crucial for deciphering complex patterns in social cybersecurity. MATLAB [246], traditionally used in engineering, extends its prowess to social network analysis, making it essential for detecting and counteracting potential cyber threats through its node metrics and ML classifiers. Python [247], versatile in its essence, empowers social cybersecurity with its vast array of libraries, such as TensorFlow, PyTorch, and NetworkX. This robust combination facilitates in-depth network analyses, predictive modeling, and intricate visualizations, offering a holistic approach to understanding and mitigating cyber vulnerabilities and threats in the realm of social networks.

**Summary**

This section provides an overview of the tools commonly utilized in social cybersecurity and social network analysis.

It covers tools for visualizing social network graphs, ML libraries like TensorFlow and PyTorch, and the Twitter API for data collection. It also highlights popular programming languages such as R, MATLAB, and Python, renowned for their extensive libraries and capabilities in social network analysis. The common approach depicted in Figure 8 encompasses the following steps in utilizing these tools:

1. *Dataset Collection:* Collect the necessary data from social media platforms, such as Twitter, by registering an application, obtaining access keys and tokens, and using APIs to retrieve relevant tweets based on social-cybersecurity keywords. Process and select a subset of the data for further analysis.

2. *Data Preprocessing:* Preprocess and clean the dataset using Python packages like NumPy, pandas, and Matplotlib or tools like AutoMap. Apply techniques such as stemming, list deletion, concept generalization, thesaurus classification, and feature selection to eliminate unnecessary text and prepare the data for analysis.

3. *Meta-Network Analysis:* Identify textual elements and link them to network nodes. Use tools like Gephi, ORA-LITE, or SocNetV to perform meta-network analysis. This involves:

   - *Network analysis:* Map the relationships that connect textual elements as a network. Identify key components (individuals and groups) and their associations to understand the community structure.

   - *Network visualization:* Transform the textual data into a visual representation that helps visualize relationships and patterns that may be difficult to discern in textual form. Visualizations aid in understanding the network structure and characteristics.

   - *Building ML models:* Utilize simulation tools like TensorFlow, Python library (PyTorch), or MATLAB to build ML models. These models can detect social cybersecurity attacks based on the analyzed meta-network data. ML algorithms can leverage the insights from network analysis to enhance detection accuracy.

By following these steps, researchers can effectively collect, preprocess, analyze, and model social media data for detecting social cybersecurity attacks. The combination of meta-network analysis, network visualization, and ML modeling provides a comprehensive approach to understanding and predicting social cybersecurity threats.

### 5.2. Public Social Cybersecurity Datasets

In this section, we present a compilation of publicly available datasets in the field of social cybersecurity. These datasets have been sourced from diverse social media platforms, including Twitter and Facebook. Each dataset is tailored to address a specific use case, encompassing tasks like friendship link prediction, rumor detection, and cyberbullying detection. To aid researchers and practitioners in their work, we also provide details regarding the feature set and size of each dataset for reference. For a succinct summary of all the discussed datasets in this section, Table 9 offers a convenient overview.

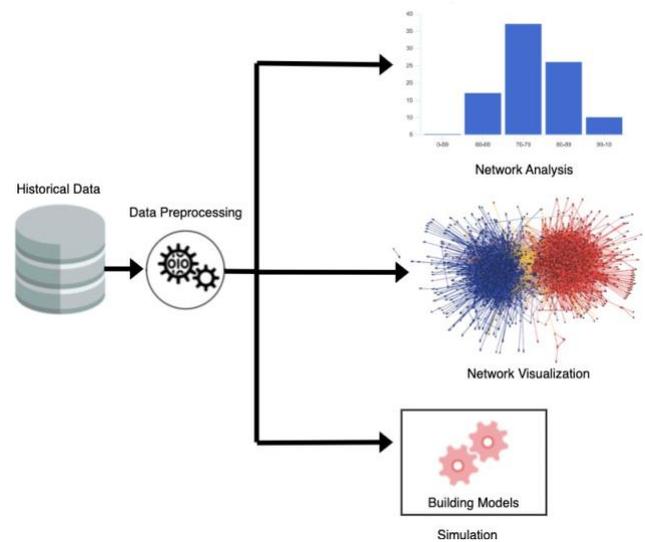

**Figure 8:** The system starts with raw data collection and ends with meta-network analysis.

In the context of Twitter datasets, several have been meticulously curated to aid various research endeavors. The Twitter dataset Social Circles [248] from the Stanford Network Analysis Project is aptly fashioned for impersonation attack detection or friendship link prediction. The rumor dataset from the PHEME Project [249] zeroes in on rumor detection. The TwiBot20 dataset [250] is inherently constructed for bot detection. The Sentiment 140 dataset curated by Stanford University [251] predominantly serves spam detection and sentiment analysis tasks. The COVID-19 Rumor dataset [252] is instrumental in detecting rumors and fake news. The CyberTweets dataset [253] is laser-focused on cyber threat detection. The How ISIS Uses Twitter dataset [254] is purpose-built for terrorism detection. The Fake-NewsNet dataset [255] emerges as a cornerstone for fake news detection tasks. Finally, the Twitter Bots Accounts dataset [256] stands out as a vital tool for differentiating between human and bot accounts.

When it comes to Facebook datasets, there are a couple of significant resources available for researchers. The Social Circles dataset from Facebook [257], a contribution from the Stanford Network Analysis Project (SNAP), is primarily tailored for impersonation attack detection and friendship link prediction. Similarly, the Facebook WOSN dataset [258], offering an undirected subgraph of user friendships, aligns well with these tasks, underscoring the essence of user relationships and interactions.

In the realm of Vulnerability datasets, social media serves as a pivotal touchstone for early detection, given the swift emergence of related discussions. Cybersecurity greatly profits from datasets tailored for vulnerability detection and severity assessment.

The ExploitDB [259] database showcases a trove of exploits instrumental for vulnerability analysis. The Microsoft Security Response Center (MSRC) [260], on the other hand, has meticulously compiled a vulnerability CVE database, divulging essential particulars concerning a multitude of vulnerabilities. In a similar vein, the CVE Details [261] database unfolds a comprehensive roster of vulnerability CVEs, facilitating granular analysis, be it vendor or product-specific, reinforcing its value for vulnerability detection.

In the category of Other datasets, cyberbullying research has seen significant contributions from a couple of pivotal datasets. The Chat Coder Formspring Cyberbullying dataset [262] from Formspring emerges as a primary tool for detecting cyberbullying, grounded in user inquiries and responses. In a similar vein, the Chat Coder MySpace Cyberbullying dataset [262] extracted from MySpace is designed to aid the pinpointing of cyberbullying events by scrutinizing user profiles and wall exchanges.

# 6. Challenges in Social Cybersecurity, Current Solutions and Future Directions

In the subsequent section, we delve into the salient challenges currently faced in the field of social cybersecurity. Despite notable progress, significant issues persist that have yet to be fully resolved. We also engage with the existing solutions to these challenges and chart the prospective pathways for future developments. Addressing these challenges is of paramount importance, as it promises to significantly reinforce the robustness and effectiveness of social cybersecurity measures. Table 10 offers a concise overview, itemizing these challenges alongside their present solutions and envisaged future trajectories.

1. **Datasets**

   The primary challenge in the domain of social cybersecurity datasets lies in the ever-evolving and often covert nature of attacks. There's a dire need for datasets that encapsulate the full spectrum of changing social engineering tactics. The clandestine nature of these attacks and the lack of standardized data collection methods pose substantial challenges, making it difficult to acquire and analyze data that would provide a holistic perspective [263; 264; 265].
   **Current Solutions:** Interdisciplinary collaborations have taken the forefront in addressing these challenges. Central to these efforts is the development of shared frameworks for data acquisition and the establishment of cohesive protocols for dataset standardization [266]. By streamlining the data collection and standardization process, and given the multifaceted nature of the data—spanning realms like social media, online discourse, and behavioral patterns—the solutions focus on harnessing advanced analytics and ensuring continuous engagement between experts in the domain.
**Future Directions:** Looking ahead, addressing the challenges posed by datasets in social cybersecurity will require innovative approaches. There is a pressing need to develop methodologies for data acquisition, standardization, and analysis that can adapt to the growing intricacies of such data [267]. Ensuring these methodologies are anchored in ethical considerations, especially around user privacy, will be crucial. By refining these techniques, the goal is to enhance the predictive capabilities of social cybersecurity models, enabling them to preemptively identify and combat threats.

2. **Absence of data-driven Metrics**

   Detecting social cybersecurity attacks introduces challenges tied to the delineation and application of suitable metrics. Conventional metrics such as accuracy or precision frequently miss the mark in capturing the multifaceted dynamics inherent in social cyber threats. This underscores the demand for domain-specific metrics sculpted through cooperative research endeavors [268; 269].
   **Current Solutions:** The quest to pinpoint a reliable ground truth for evaluation, considering the human-centric and psychological elements woven into these attacks, necessitates a multifaceted approach. A potent solution is rooted in data sharing and standardized benchmarking. As prediction paradigms become more intricate, there's an amplified call for metrics that can be effortlessly interpreted and expounded upon. Leveraging tools such as explainable AI and rule-based reasoning have emerged as viable routes to amplify the interpretability quotient of these metrics [270].
   **Future Directions:** There's an imminent need to continue refining and amplifying the robustness of metrics tailored for social cybersecurity. Emphasis should pivot toward solidifying evaluation techniques that marry precision with practical relevance. A sustained focus on ensuring that metrics are not just numerically rigorous but also intuitively comprehensible will be paramount. Collaborative interdisciplinary research should surge ahead, seeking to decode the complexities of human behaviors in cyber settings [271].

3. **Time Complexity of Various Techniques**

   Efficiently grappling with the time complexity inherent in techniques designed for detecting social cybersecurity attacks poses considerable challenges. As the sheer volume and velocity of data intensify, there exists a compelling need for prediction models that excel in both efficiency and scalability. The intricate nature of social cybersecurity attacks demands swift, real-time analysis, presenting substantial hurdles for researchers and practitioners [272; 273].

**Current Solutions:** The rapidly evolving landscape of social cybersecurity attacks, accompanied by overwhelming data quantities, necessitates continual advances in algorithm design, optimization paradigms, and parallel computing. To strike a balance between computational precision and efficiency, the focus has shifted towards the efficient design of algorithms and data handling mechanisms [274]. Techniques such as feature selection, dimensionality reduction, and tapping into distributed and cloud-based computing infrastructures have become quintessential. The aim is to design algorithms that not only handle vast datasets but also scale efficiently, supported by computational platforms capable of parallel processing and efficient data distribution.

**Future Directions:** The future will demand further refinement of the delicate balance between algorithmic efficiency and scalability. As cyber threats continue to evolve, there's a pressing need for algorithms that remain efficient under increasingly complex conditions. There will be a pivot towards leveraging quantum computing, advancements in machine learning, and innovative data representation techniques. Interdisciplinary collaborations will be vital, focusing on solutions that are both reactive and proactive, fortifying our defenses against the relentless evolution of social engineering attacks [275].

4. **Evolving Attack Methods and Human Behavior's Complexity**

   The landscape of social cybersecurity is perpetually challenged by two interlinked dimensions: the rapid evolution of attack methodologies and the intricate puzzle of human behavior. The dynamic nature of technological advancements and human psychology complicates the prediction and detection of social cybersecurity threats. Attackers not only exploit new communication technologies but also human vulnerabilities, making the forecasting of their next moves increasingly complex [276; 277; 278; 279; 280; 281].

   **Current Solutions:** Addressing these challenges necessitates a multifaceted approach that combines vigilant monitoring of evolving attack strategies with an in-depth understanding of human behavior. This includes incorporating psychological, behavioral, and sociological insights into predictive models, alongside real-time threat intelligence. Collaborative ecosystems, uniting academia, cybersecurity experts, and industry leaders, play a crucial role in fostering an adaptable and informed response mechanism. Bridging diverse academic disciplines and amassing culturally rich behavioral datasets, while upholding privacy and ethical standards, are essential for developing effective countermeasures [282; 283].

   **Future Directions:** The future of social cybersecurity hinges on enhancing the detection accuracy of models through the integration of machine learning, AI, and culturally intelligent analytics. These models must be adaptive, capable of deciphering complex human behaviors and evolving technological contexts. Emphasizing ethical AI and robust data governance will ensure that advancements in cybersecurity respect human dignity and privacy. As we navigate this ever-changing domain, the synergy between cutting-edge technology and deep human behavioral insights, underpinned by unwavering ethical principles, will be paramount in securing a safer digital world [284; 285].

5. **Adapting to Social Media Changes**

   The perpetual evolution of the social media landscape presents a significant challenge in accurately detecting social cybersecurity attacks [286; 287]. As attackers leverage novel features and emerging communication channels, identifying vulnerabilities and potential attack vectors across numerous platforms becomes increasingly complex.

   **Current Solutions:** The current efforts predominantly revolve around continuous monitoring of social media platforms to detect anomalies and potential threats. Collaboration is also underway between researchers, cybersecurity experts, and platform administrators to share insights and best practices. To keep pace with the speed of information dissemination on social media and the tactics used by adversaries, there is a reliance on advanced threat intelligence tools and cutting-edge analytics. This approach aims to ensure that the most up-to-date and relevant data is used to identify and counter threats in real time [288].

   **Future Directions:** As the social media terrain becomes even more intricate, there will be an imperative for developing adaptive modeling frameworks. These frameworks will need to dynamically adjust to the intricacies of new communication channels and features on various platforms. The next phase of defense will focus heavily on leveraging AI and machine learning, not just for detection but for proactive threat prediction. Anticipating an attacker's move before they make it will become crucial. Additionally, building a robust and more integrated collaborative network, which includes not only security experts but also social media platform developers, will be pivotal. This collaboration will aim to introduce security measures at the very design level of social platforms, ensuring that security becomes an integral part of the social media evolution [289; 290].

Table 8: A summary of the social cybersecurity tools surveyed.

| Ref. | Tool | Purpose | Type | Platform | Open Source | Language | Input Formats | Output Formats |
|------|------|---------|------|----------|-------------|----------|---------------|----------------|
| [227] | Pajek | Visualization, analysis of large networks | Desktop App | Windows | X | .NET | .net, .pajek, .dat, .dl, .gml, .gdf, .csv, .mol, .xls | .eps, .svg, .html, .jpeg, .bmp, .X3D, .KiNG, .mdl |
| [235] | NetworkX | Visualization, analysis of networks & Graphs | Library | Windows, Linux | ✓ | Python | .gml, .graphml, networkx,graph6, sparse6, .dot, .net,.gexf, .txt | .gml, .ps, .dot, graph6/sparse6, .net adjacency lists, edge lists, .jpg, .png, .graphml |
| [236] | igraph | Creation, analysis of graphs | Library | Windows, Linux | ✓ | C | .net, .graphml, .gml, .text, .csv, .dot, Graph db, .txt | .net, GraphML |
| [228] | Gephi | Graph manipulation | Desktop App | Windows, Linux, Mac | ✓ | Java | .net, .graphml, .net, .gml, .gml, .vna, .xls, .gdf, .dot, .txt, .csv, .tlp, .dl, .tpl, .gexf, .vna | .net, .dl, .gexf, .gdf, node lists, edge lists, .graphml |
| [237] | JUNG | Manipulation, visualization, analysis of graphs | Library, API | Windows, Linux, Mac | ✓ | Java | Pajek and GraphML, .txt | Pajek, GraphML |
| [229] | AutoMap | Text mining | Desktop App | Windows | X | Java | .txt | DyNetML, .csv |
| [231] | UCINET | Visualization, analysis of social networks | Desktop App | Windows | X | BASIC/ DOS | .dl, .xls, vna, .net, .txt | dl, .xls, .net, Mage, Metis, Netdraw (.net) |
| [232] | SocNetV | Visualization, analysis of social networks | Desktop App | Windows, Linux, Mac | ✓ | C++ | .graphml, .xml, .dot, .net, .paj, .sm, .csv, .adj, .dl, .list, weighted lists (.wlist) | GraphML, PDF, Pajek, .jpeg, .png, Adjacency matrix |
| [233] | CFinder | Discovery, visualization of communities | Desktop App | Windows, Linux, Mac | ✓ | Java | .txt | .txt, .pfd, .gif, .jpg, .png, .bmp, .ps, .wbmp, .svg, .emf |
| [238] | NodeXL | Discovery, visualization of networks | Plugin | Windows | ✓ | .NET C# | .txt, .csv, .net, .xls, .xslt, .dl, .graphml | .txt, .csv, .dl, .xls, .xslt, .graphml |
| [234] | Hootsuite | Management of social media | Desktop, Mobile App | Windows, Linux, IOS, & Android | X | PHP | N/A | N/A |
| [239] | JGraphT | Graph illustration | Library | Windows, Linux, Mac, & Android | ✓ | C#, Java, Javascript | N/A | .dot, .txt |
| [245] | R | Analysis of social networks | Language | Windows, Linux, Mac | ✓ | C++, C, Fortran | .R, .RData, .rds, .rda | .txt, csv, .html, .xml |
| [246] | MATLAB | Analysis of networks, numeric operations | Language | Windows, Linux, Mac | X | C, C++, Java | .mat, .txt, .csv, .xls, .xltm, .ods, .xml, .daq, .cdf, .avi, .hdf, .bmp, .jpg, .png, .pbm, .pcx, .tiff, .au, .wav, .mpg | .mat, .csv, .xls, .xltm, .ods, .xml, .cdf, .fits, .hdf, .bmp, .jpg, .png, .pbm, .pcx, .tiff, .au, .wav, .mpg, .mp4, .avi |
| [247] | Python | Visualization, analysis of social networks, text processing, ML | Library | Windows, Linux | ✓ | C | csv, .xslx, .txt, .json, .docx, .jpeg, .mp3, .mp4, .sql | csv, .xslx, .txt, .json, .docx, .jpeg, .mp3, .mp4, .sql |
| [240] | Twitter API | Collect Twitter datasets | API, Library | Windows, Linux, Mac, IOS, Android | ✓ | Java, Javascript | N/A | .json |
| [241] | Tensorflow | Creation of ML classifiers | Library | Windows, Linux, Mac | ✓ | Python, C++, Cuda | .pb, .json | .pb, .json |
| [242] | PyTorch | Creation of ML classifiers | Library | Windows, Linux, Mac | ✓ | Python, C++, Cuda | .pt, .pth | .pt, .pth |
| [243] | MuxViz | Visualization and analysis of multilayered networks | Library, Website | Windows, Linux, Mac | ✓ | R | .dot, .paj, .txt, .json, .csv | .net, .dl, .net, .gdf, .xlsx, .png, .txt, .json, .csv |

**Table 9**
A summary of the public datasets surveyed.

| Ref. | Name | Source/Type | Size | Features | Use Cases |
|---|---|---|---|---|---|
| [248] | SNAP: Social circles on Twitter | Twitter | 81,306 nodes and 1,768,149 edges | Various hashtags and mentions | Impersonation attack detection and friendship link prediction |
| [249] | PHEME rumor dataset | Twitter | 330 threads (297 in English, 33 in German) associated with nine breaking news stories | Thread features, Tweet features, and replying Tweet features | Rumor detection |
| [250] | TwiBot20 | Twitter | 229,573 users, 33,488,192 tweets, 8,723,736 user properties, and 455,958 follow relationships | User features, neighbors, domain, and label | Bot detection |
| [251] | Sentiment140 | Twitter | 498 annotated tweets and 1.6 million processed tweets | Tweet features and polarity | Spam detection and sentiment analysis |
| [252] | COVID-19 Rumor Dataset | Twitter and Web news | 4,129 news records and 2,705 tweets with replies | Tweet features and news features | Rumor detection and fake news detection |
| [253] | CyberTweets | Twitter | 21,368 tweets | Tweet features, relevance, and vulnerability type | Cyber threat detection |
| [254] | How ISIS uses Twitter | Twitter | 17k tweets from 112 pro-ISIS accounts | User features and tweet features | Terrorism detection |
| [255] | FakeNewsNet | Twitter | Real-time growth | News article features, Tweet features, and user features | Fake news detection |
| [256] | Twitter Bots Accounts | Twitter | 25,013 human accounts and 12,425 bot accounts | Twitter user ID and label | Bot detection |
| [257] | SNAP: Social Circles on Facebook | Facebook | 4,039 nodes and 88,234 edges | User features | Impersonation attack detection and friendship link prediction |
| [258] | Facebook WOSN | Facebook | 63,731 nodes in an undirected subgraph of friendships | flraph features | Impersonation attack detection and friendship link prediction |
| [259] | ExploitDB | vulnerability dataset | N/A | CVE features | Vulnerability detection and severity rating |
| [260] | Microsoft Security Response Center (MSRC) | vulnerability dataset | N/A | CVE features | Vulnerability detection and severity rating |
| [261] | CVE Details | vulnerability dataset | N/A | CVE features | Vulnerability detection and severity rating |
| [262] | Chat Coder Formspring Cyberbulling Dataset | Formspring | 18,554 users with questions and answers | User features and post features | Cyberbullying detection |
| [262] | Chat Coder MySpace Cyberbulling Dataset | MySpace | Profiles and walls of 127,974 users | User features and post features | Cyberbullying detection |
| [291] | 2018 Master Table - Hackmageddon | Various sources | 1,337 events | Attack features | Cybercrime detection and prediction |
| [292] | The fIDELT Project | Various sources | Real-time growth | Event codes and locational features | Prediction of social unrest |

## 6. Information Maneuvers in Social Cyber Attacks

The realm of information maneuvers in social cyber attacks remains an intricate landscape [6; 293; 294]. These maneuvers, both nuanced and multifaceted, highlight the deliberate strategies cyber adversaries deploy to achieve specific outcomes or impacts. The challenge lies in enhancing detection capacities to swiftly and proactively pinpoint these information maneuvers, especially when they span across multiple digital platforms. The ultimate objective is to build a more resilient digital space fortified against the crafty ploys of cyber attackers.

**Current Solutions:** In the bid to counteract these meticulously orchestrated maneuvers, contemporary solutions have hinged on the zenith of technological innovation. Foremost among them are real-time monitoring systems. Bolstered by machine learning and big data paradigms, these systems perpetually trawl through digital content, isolating aberrant patterns indicative of information maneuvers [295]. Adding depth to these solutions are sentiment and intent analysis algorithms. Beyond mere content analysis, they delve into the deeper nuances, aiming to discern the actual sentiment or underlying intent, particularly when confronted with content that bears the hallmarks of manipulation or subterfuge [296]. Given the omnipresence of digital platforms, the introduction of cross-platform correlation tools has been a game-changer. These tools, adept at synthesizing data and behavior across myriad social media channels, provide a holistic threat overview, ensuring comprehensive vigilance [297].

**Future Directions:** As cyber attackers refine their information maneuvers, the counter-strategies are bound to evolve in tandem. Poised on the horizon are adaptive analytics systems. Anchored in AI, these systems will incessantly adapt, always staying a step ahead of the ever-mutating tactics of cyber adversaries. With information warfare witnessing incessant sophistication, the corresponding counter-strategies will pursue heightened predictive acuity and instantaneous responsiveness. The fusion of behavioral psychology with cybersecurity presents an intriguing prospect. Through a more profound understanding of the psychological triggers attackers leverage, the potential to proactively stymie their maneuvers becomes increasingly tangible. As the tapestry of the digital universe gets more intricate, collaboration will emerge as the clarion call. United defense stratagems, spanning platforms, nations, and sectors, are projected to be the future bulwark against the ever-evolving realm of information maneuvers [298].

7. **Motive Identification in Social Cybersecurity Attacks**

Understanding the driving forces behind social cybersecurity attacks is paramount [299; 300; 301]. The motives, be they rooted in personal amusement, quest for chaos, financial objectives, brand promotion, wielding personal influence, or community creation, remain as diverse as they are intricate. Given the scale and rapid evolution of cyber threats, the challenge rests on effectively discerning these motives in real-time. The pressing need, thus, is to cultivate innovative methodologies that can delve deep and provide actionable insights into the motivations underpinning these cyber onslaughts.

**Current Solutions:** The contemporary approach to grasping the psyche of cyber adversaries marries technological prowess with psychological acumen. Machine learning, specifically in the realm of behavior analytics, stands out as a primary tool to glean patterns potentially indicative of a cybercriminal's motivations [302]. The application of NLP has carved a niche, especially when malefactors articulate their objectives or leave behind textual breadcrumbs, thereby streamlining the motive identification process [303]. In a bid to foster global collaboration, platforms have been inaugurated that rally cybersecurity mavens worldwide. This collaborative spirit has culminated in the creation of robust motive repositories, expediting the process of associating discerned behaviors with probable motivations.

**Future Directions:** The realm of motive identification promises an exciting trajectory. Foreseen is the amalgamation of AI's behavioral analytics with the intricacies of deep learning, offering a refined lens to scrutinize motives [304]. As cyber-attacks manifest in multifarious shades, a converging path with disciplines such as sociology and psychology seems inevitable, laying the foundation for a more comprehensive grasp of motivations. Innovations like real-time motive detection systems, underpinned by edge computing and IoT integration, are on the horizon, ensuring instant motive discernment during live cyber events. In this hyper-connected digital epoch, international synergies are more pertinent than ever. The emergence of standardized cyber platforms and consolidated motive databanks is anticipated, ensuring a seamless, global collaborative effort to comprehend and neutralize cyber threats anchored in clear motive understanding.

8. **Diffusion in Social Cybersecurity Attacks**

Understanding the diffusion mechanisms in social cybersecurity attacks poses a multi-faceted challenge [305; 306; 290]. The crux of this research lies in capturing the spread of influence campaigns, which are often multi-modal, encompassing a diverse range of content from memes and videos to beliefs and ideas. Essential to this is the ability to trace the originators of these campaigns and gauge the cascading effects as they resonate across various platforms. There is a distinct need for innovative methodologies and live-monitoring systems that can adeptly detect diffusion trajectories, spanning initiation, peak momentum, and eventual decline.

**Current Solutions:** The increasing prevalence of influence campaigns has catalyzed the emergence of state-of-the-art diffusion analysis tools. Pioneering analytics platforms now offer real-time tracking of content trajectories across social media ecosystems, buoyed by sophisticated AI algorithms [307]. The insights gleaned from network analysis have been invaluable, shedding light on diffusion pathways and pinpointing influential nodes or super-spreaders [308]. Sentiment analysis tools harnessing the capabilities of NLP provide a nuanced understanding of how content is received, guiding adaptive strategies [309]. In response to the intricate web of interconnected social platforms, cross-platform analytics have gained prominence, offering a consolidated lens to view and comprehend diffusion dynamics.

**Future Directions:** As the digital sphere continues to expand and metamorphose, so will the paradigms of diffusion research in social cybersecurity. The horizon likely holds promise for quantum-powered real-time analytics, poised to revolutionize data processing speeds and accuracy [310]. The melding of Augmented Reality (AR) and Virtual Reality (VR) with diffusion analytics is anticipated, paving the way for immersive and intuitive exploration of diffusion patterns. Advanced deep learning models capable of discerning and forecasting content virality based on past patterns are on the cards [311]. With the mounting sophistication of cross-platform influence campaigns, global collaborative endeavors are expected to take center stage. This could manifest as unified platforms that empower cybersecurity specialists worldwide to collaboratively track, decipher, and counter diffusion mechanisms, ensuring a cohesive defense against proliferating cyber adversities.

9. **Authenticity of Open-Source Code in Social Cybersecurity Attacks**

The challenge of ensuring the authenticity of open-source code has become a pressing concern in the realm of social cybersecurity [312; 313; 314; 315]. With an increasing dependence on open-source software in numerous sectors, the integrity and authenticity of such code have become paramount. Malicious actors, recognizing this reliance, have attempted to compromise systems by introducing backdoors, injecting malicious code, or presenting deceptive code repositories. This presents a dual problem: the need to ensure that open-source contributions are genuine and free from harmful components and the challenge of detecting and countering attempts at misinformation through code.

**Current Solutions:** Several solutions focus on ensuring code security. Code signing allows developers to attach digital signatures to their code, which users can verify before execution [316]. Tools like SAST and DAST are employed to identify potential vulnerabilities [317]. Software Composition Analysis (SCA) tools trace open-source components, ensuring they are up-to-date and free of known vulnerabilities [318].

**Future Directions:** Machine learning and AI could be harnessed for in-depth code analysis, identifying malicious code patterns [319]. Blockchain technologies suggest the potential creation of decentralized code repositories, promoting transparent code modification documentation [320]. Collaborative platforms emphasizing peer reviews and ratings could ensure secure code contributions. Emphasis might also be on educational initiatives, equipping developers with knowledge on secure coding practices. A collaborative international approach could lead to the establishment of global standards for open-source software.

10. **Ethical Implications of AI-based Cybersecurity**

A significant challenge often overlooked in social cybersecurity is the ethical dimension, particularly regarding AI-based detection and monitoring systems. The application of AI in this domain often involves extensive data collection, monitoring, and profiling of users, which raises serious concerns about user privacy, digital rights, consent, and potential misuse of collected data [321]. These practices can lead to inadvertent surveillance and discrimination, potentially violating ethical norms and user expectations.

**Current Solutions:** Recent studies advocate for the integration of ethical AI frameworks that incorporate fairness, accountability, transparency, and explainability into social cybersecurity systems [322]. Privacy-preserving techniques, such as federated learning and differential privacy, are increasingly being explored to reduce the amount of personally identifiable information collected during monitoring activities [323].

**Future Directions:** Future research should aim at designing AI-driven social cybersecurity tools that balance the need for security with fundamental ethical principles. There is a pressing need to establish regulatory frameworks and standardized ethical guidelines to ensure responsible AI use [324]. Addressing issues such as bias mitigation, algorithmic accountability, and user consent will be essential to gaining public trust and maintaining the legitimacy of AI-based social cybersecurity measures.

## 7. Conclusion and Future Directions

This research paper offers a thorough exploration of detection in the context of social cybersecurity attacks. We systematically dissected a range of social cybersecurity attacks, providing an in-depth analysis of potential countermeasures for each, giving readers a comprehensive understanding of the field. A key highlight of our paper is the emphasis on the importance of public datasets and specialized analytical tools for social cybersecurity. Such resources not only enhance the quality of research but also catalyze advancements in the discipline.

Furthermore, we critically examined the vast array of existing detection techniques. We brought to the fore the research challenges in social cybersecurity, ongoing solutions, and potential future pathways. Our discourse illuminated the intricate challenges of countering social engineering threats and underscored the urgent need for sustained innovation in the sector.

Looking forward, it's evident that the next phase in social cybersecurity requires delving into unexplored areas and pioneering innovative research directions. By taking head-on the challenges we've identified, and by adopting cutting-edge approaches, the cybersecurity community is poised to bolster its defenses against the ever-evolving threats of malicious online actors.

This survey also highlights key policy implications and actionable recommendations for practitioners:

- Policy Implications: Governments and organizations should prioritize the development and implementation of regulatory frameworks that address social cybersecurity threats, ensuring platform accountability and fostering user trust.

- Public Awareness Campaigns: Investing in educational initiatives to raise public awareness about social engineering tactics, phishing attempts, and other threats is essential in building a resilient digital society.

- Technological Investments: Practitioners should focus on deploying advanced AI-driven detection algorithms and multi-layered authentication mechanisms to proactively counter cybersecurity threats.

- Collaboration and Intelligence Sharing: Establishing cross-platform collaborations and intelligence-sharing networks among stakeholders is critical for identifying, predicting, and mitigating emerging threats.

In essence, this survey serves as a guiding beacon for researchers, professionals, and decision-makers immersed in social cybersecurity. Our objective has been to shed light on the ever-changing landscape of this domain, driving the design of advanced detection models, enhancing defensive measures, and nurturing a more secure digital space. Through the insights presented in this paper, we are rallying our collective efforts, aiming for a robust and resilient digital tomorrow. As we venture forward, it's apparent that the field of social cybersecurity is ripe for breakthroughs and daring exploration, especially in devising effective countermeasures against complex social engineering threats.

**Table 10**
Summary of challenges, evaluations, current solutions, and potential solutions in social cybersecurity.

| Challenge | Evaluation | Current Solutions | Future Directions |
|---|---|---|---|
| Datasets | The primary challenge in the domain of social cybersecurity datasets lies in the covert nature of attacks [263; 264; 265]. | Interdisciplinary collaborations focus on shared frameworks for data acquisition [266]. | Innovative approaches are needed for data acquisition, standardization, and analysis [267]. |
| Absence of data-driven Metrics | Challenges tied to suitable metrics for forecasting social cybersecurity attacks [268; 269]. | Data sharing and standardized benchmarking are essential. Use of tools like explainable AI [270]. | Refining and amplifying metrics tailored for social cybersecurity is essential [271]. |
| Time Complexity of Various Techniques | Time complexity challenges in techniques for detecting attacks [272; 273]. | Focus on algorithm design, optimization paradigms, and parallel computing [274]. | Emphasis on leveraging quantum computing and machine learning advancements [275]. |
| Evolving Attack Methods and Human Behavior's Complexity | The interplay between the rapid evolution of attack methodologies that leverage new technological breakthroughs and the complexity of human behavior, including social engineering attacks and cultural variances, presents a dual challenge [276; 277; 278; 279; 280; 281]. | A multifaceted strategy that includes vigilant monitoring of evolving attack strategies, real-time threat intelligence, an interdisciplinary approach that melds psychology, behavioral sciences, and sociological insights, and fostering collaborative ecosystems [282; 283]. | Embracing machine learning, AI, and heuristic-driven analytics, enhanced with deep insights into human psychology, to develop adaptive, robust, and ethically guided predictive models that can navigate both the technological and human-centric aspects of cybersecurity threats [284; 285]. |
| Adapting to Social Media Changes | Evolution of social media challenges accurate attack detection [286; 287]. | Continuous monitoring, collaborations, and advanced threat intelligence tools [288]. | Adaptive modeling frameworks, leveraging AI for proactive detection, and integrated collaborations [289; 290]. |
| Information Maneuvers in Social Cyber Attacks | Challenging, especially when spanning multiple platforms [6; 293; 294]. | Real-time monitoring and sentiment/intent analysis algorithms [295; 296; 297]. | Adaptive analytics systems and cross-platform, cross-nation collaborations [298]. |
| Motive Identification in Social Cybersecurity Attacks | Difficult due to diverse and rapidly evolving threats [299; 300; 301]. | ML behavior analytics and NLP for textual analysis [302; 303]. | Integration of AI with deep learning and real-time motive detection systems [304]. |
| Diffusion in Social Cybersecurity Attacks | Multi-faceted challenge in understanding spread of influence campaigns, which are multi-modal and diverse [305; 306; 290]. | State-of-the-art diffusion analysis tools with real-time tracking, network analysis, sentiment analysis tools, and cross-platform analytics [307; 308; 309]. | Quantum-powered real-time analytics, integration of AR/VR with diffusion analytics, advanced deep learning models, and global collaborative endeavors [310; 311]. |
| Authenticity of Open-Source Code in Social Cybersecurity Attacks | Rising dependence on open-source software, need to ensure code's integrity, and challenges from misinformation [312; 313; 314; 315]. | Code signing, SAST and DAST tools, and Software Composition Analysis tools [316; 317; 318]. | ML and AI for code analysis, blockchain for decentralized repositories, peer reviews, and global standards for open-source [319; 320]. |
| Ethical Implications of AI-based Cybersecurity | Use of AI in social cybersecurity raises concerns about user privacy, digital rights, potential bias, and unintentional surveillance due to extensive data collection and monitoring [321; 322]. | Integration of ethical AI frameworks emphasizing fairness, accountability, transparency, privacy-preserving techniques (e.g., federated learning, differential privacy) [323]. | Establishing regulatory frameworks and standardized ethical guidelines, incorporating explainable AI, and fostering public trust through responsible AI practices [324]. |